\documentclass[a4paper,12pt]{article}
\usepackage{CJK}
\usepackage{color}
\usepackage{graphicx}
\usepackage{epstopdf}
\usepackage{amssymb}
\usepackage{amsfonts}
\usepackage{amsmath}
\usepackage{mathrsfs}
\usepackage[numbers,sort&compress]{natbib}
\usepackage{epsf}
\usepackage{bm}
\long\def\begincomment#1\endcomment{}
\usepackage{titletoc}

\usepackage{xcolor} 
\newcommand{\bt}[1]{{\color{blue} #1}}

\begin{document}
\setlength{\baselineskip}{7mm}

\title{\textbf{Phase Synchronization on Spacially Embeded Duplex Networks with Total Cost Constraint}}
\author{ Ruiwu Niu \footnotemark[1],
         Xiaoqun Wu \footnotemark[1] $^,$\footnotemark[3],
         Jun-an Lu \footnotemark[1],
         Jianwen Feng \footnotemark[2]}
\renewcommand{\thefootnote}{\fnsymbol{footnote}}
 \footnotetext[1]{School of Mathematics and Statistics, Wuhan University, Wuhan 430072, China.}
 \footnotetext[2]{College of Mathematics and Statistics, Shenzhen University, Shenzhen 518060, PR China}
 \footnotetext[3]{To whom correspondence should be addressed: xqwu@whu.edu.cn}
 \footnotetext[4]{This work was supported in part by the National Natural Science Foundation of China under Grants 61573262.}

\date{}
\maketitle{}

\begin{abstract}
Synchronization on multiplex networks have attracted increasing  attention in the past few years. We investigate collective behaviors of Kuramoto oscillators on single layer and duplex spacial networks with total cost restriction, which was introduced by Li et. al [Li G., Reis S. D., Moreira A. A., Havlin S., Stanley H. E.  and Jr A. J., {\it Phys. Rev. Lett.}  104, 018701 (2010)] and termed as the Li network afterwards. In the Li network model, with the increase of its spacial exponent, the network's structure will vary from the random type to the small-world one, and finally to the  regular lattice.
We first explore how the spacial exponent   influences the synchronizability of Kuramoto oscillators on single layer Li networks and find that   the   closer the Li network  is to a regular lattice, the more difficult for it to evolve into synchronization.   Then we investigate synchronizability of   duplex Li networks and find that the existence of inter-layer interaction can greatly enhance inter-layer and global synchronizability. When the inter-layer coupling strength is larger than a certain critical value, whatever the intra-layer coupling strength is, the inter-layer synchronization will always occur. Furthermore,  on single layer Li networks,  nodes with larger degrees more easily  reach global synchronization, while on duplex Li networks, this phenomenon  becomes much less obvious. Finally,
we study the impact of  inter-link density on global synchronization  and obtain that  sparse inter-links can lead to the emergence of global synchronization for duplex Li networks just as dense inter-links do.  In a word, inter-layer interaction plays a vital role in determining  synchronizability for duplex spacial networks with total cost constraint.

\end{abstract}



\section{\label{sec:level1}Introduction}

Synchronization as a natural phenomenon, which can be easily observed in a wide variety of biological, chemical and physical systems, has drawn extensive attention of scientists in the last few decades. A large amount of research achievements have been made about synchronization properties of small-world, scale-free and other types of complex networks~\cite{WS1998,Barahona2002,Hong2002,Newman2003,Lu2004,Lu2005,Zhou2006,Boccalettia2006,Arenas2008,ChenY2011,Huang2008,
Donetti2005,Pecora2014, TangLK2012a,Majun2011,Majun2017,Fengjw2017}. Aside from these researches, the spacial network has become more and more popular ever since it was proposed in 2000~\cite{Kleinberg2000}. As we know, many natural or man-made infrastructures are embedded in space, such as power grids, oil pipelines, communication networks and so on. So, it is important to investigate the synchronization properties of spacially embedded networks.

Among various studies on  network synchronization,  the Kuramoto model has been extensively employed~\cite{Kuramoto1984}. It can describe collective behaviors caused by interaction between coupled  oscillators. Since then, many works have been done to analyze synchronization among coupled oscillators. Watts and Strogatz found that a few shortcuts between  Kuramoto oscillators in a network can greatly improve the synchronizability of the network~\cite{Watts1999,Strogatz2001}. Moreno and Pacheco applied   Kuramoto oscillators to scale-free networks and indicated that the hubs play a vital role in determining network dynamics~\cite{Moreno2004}. After that, some significant analytical and numerical results demonstrated the relation between synchronization and network topologies~\cite{Boccalettia2006,Arenas2008}. In 2011, G\'{o}mez et. al discovered a  discontinuous synchronization transition  by introducing a correlation between network structure and local dynamics~\cite{GG2011}. This special phenomenon called explosive synchronization offers a new perspective  for  investigating synchronization of the Kuramoto model.

It is noteworthy that many synchronization phenomena, as in social networks, do not involve a single network in isolation but rely on the behaviors of a collection of smaller networks~\cite{Boccaletti2014,Gregorio2014}. And more generally, beyond single networks, we are now understanding that interactions between networks are playing an increasingly important role  in determining  the dynamical processes~\cite{PNAS2012,Radicchi2013,Gomez2013,Domenico2014,Valles-Catala2016}.

 In the last few years, a new type of networks called multilayer networks has been paid increasing attention to.  Numerous results indicate that  interaction between networks can greatly affect dynamics on  interacting networks,  such as spreading,  diffusion  and synchronization~\cite{Ribalta2013,Aguirre2014,Boccaletti2014,Demomenico2015, LuRQ2014,Luo2014,Liyang2015,Xumm2015,
Gambuzza2015,Sevilla2016,Wang2016,Liu2016,Mei2017}. For example, in 2014,  Aguirre et al. \cite{Aguirre2014}  investigated
the  influence of the connector node degree on the synchronizability of two star networks
with one inter-layer link and showed that connecting the high-degree (low-degree) nodes
of each network is the most (least) effective way to achieve synchronization. In 2016, Wei et al.~\cite{Wei2016}worked on cooperative epidemic spreading on two interacting networks and found that   the global epidemic threshold    is smaller than the epidemic thresholds of the corresponding isolated networks and the cooperative interaction between networks can enhance the final infection fraction.
Very recently, Tang et. al~\cite{Tang2016} studied   inter-layer, intra-layer and global synchronization in two-layer networks, and showed that for any give nodal dynamics and network structure, the occurrence of intra-layer and inter-layer synchronization depend mainly on the coupling functions of nodes within a layer and across layers, respectively.

In the real world, many complex networks can be geographically represented or spatially embedded. In 2000~\cite{Kleinberg2000}, Kleinberg proposed a spatial network model by adding shortcuts with probability \(P(r_{ij})\sim r^{-\alpha}_{ij}\) on a regular lattice, where \(r_{ij}\) is the Manhattan distance between nodes \(i\) and \(j\) on the lattice, and \(\alpha\) is the spacial  exponent. He found that proper addition of shortcuts (\(\alpha=d\), where \(d\) is the spatial dimension) can lead to   formation of  a complex network possessing the small-world property.  After that, considering the cost of connecting different nodes, Li et. al improved Kleinberg's model by restraining the total length of shortcuts, which is named as Li network afterwards~\cite{Li2010,Li2013}.  They found that when \(\alpha=d+1\) , the network has the small-world property.  These results have inspired a lot of   studies on this interesting phenomenon of the emergence of the small-world property on lattices by   shortcuts addition ~\cite{Hu2011,Oliveira2014,Pan2016,Niu2016}. Nevertheless, few of them focused on synchronous behaviors in spatial networks. In this paper, we  consider the specific  Li network of Kuramoto oscillators and find that
   proper addition of shortcuts can greatly improve the synchronizability of spacial networks. 

In order to investigate synchronous behaviors on mutilayer spatial networks, we propose a spatially embedded duplex network model with total cost constraint. Here, the same set of nodes interact on different layers, and links in different layers represent different link types.  By employing the Kuramoto model, we study inter-layer, intra-layer and global synchronizability of the networks. We show how  coupling strength between layers influences synchronization, and point out that large inter-layer coupling strength can greatly enhance the inter-layer, intra-layer as well as global synchronization. We further investigate the synchronization processes. Interestingly, we find that, even for a small value of inter-layer coupling strength, inter-layer synchronization can always happen in the first place. In addition, for single layer networks, the nodes with large degrees always firstly arrive at synchronization.  While for duplex networks, this phenomenon becomes less obvious due to inter-layer connections. At last, we reduce the inter-links between layers and find that, a small portion of inter-links can make the networks achieve global synchronization just as one-to-one inter-layer connections can do.

\section{\label{sec:level1}Results and Discussion}

\subsection{\label{sec:level2}Synchronization on single layer Li network}
First of all, we introduce a spatial network proposed by Li et. al~\cite{Li2010,Li2013}. Figure 1 shows a typical Li network, which is a regular two-dimensional square lattice with \(N=L\times L\) nodes, where \(L\) is the linear size of the lattice, and long-range connections are randomly added between nodes. In this model, pairs of nodes \(i\) and \(j\) are randomly chosen to generate a long-range connection with probability \(P(r_{ij})\sim r^{-\alpha}_{ij}\) , where \(r_{ij}\) is the Manhattan distance between nodes \(i\) and \(j\), and \(\alpha\) is the spacial exponent. In addition, the total length of the long-range connections is restricted by \(\wedge=L\times L\) . It is noteworthy that if exponent \(\alpha=d+1\) , the network will have the small-world property~\cite{Li2010,Li2013}. Besides, when \(\alpha\) is getting larger, the network will be closer to a regular lattice. Otherwise, the network will become a lattice network with few long shortcuts. In this way, we can change network topology by altering the exponent \(\alpha\).

\begin{figure}[!ht]
  \centering
 \includegraphics[width=8cm]{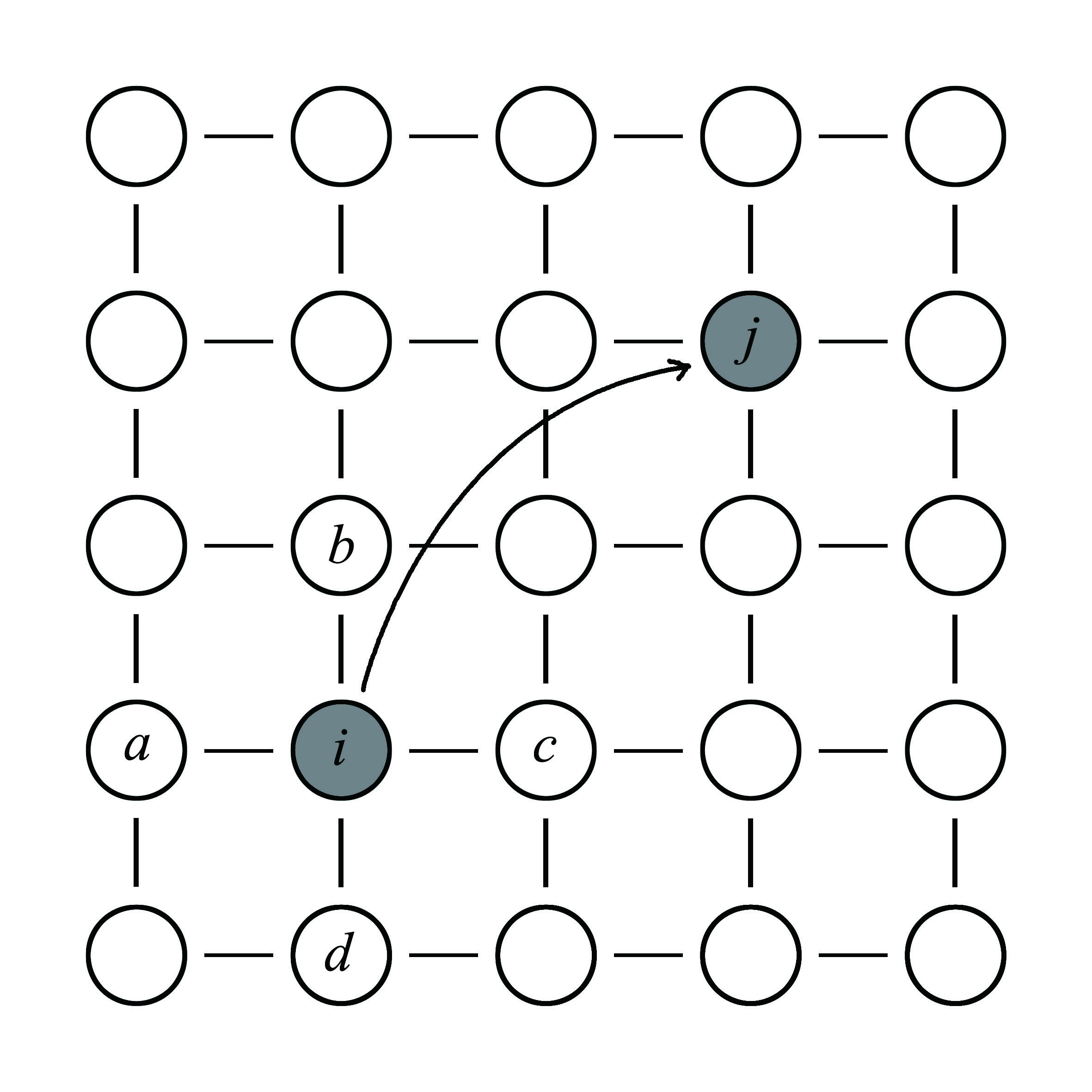}\\
     \caption{In a two-dimensional space, each node \(i\) has four short-range connections to its nearest neighbors (\(a\),\(b\),\(c\) and \(d\)). A long-range connection is added to a randomly chosen node \(j\) with probability proportional to \(r_{ij}^{-\alpha}\).}\label{single layer}
\end{figure}

Next, consider the Kuramoto oscillators as node dynamics. Thus the evolution of the \(N\) oscillators is governed by~\cite{Kuramoto1984}

\begin{equation}\label{kuramoto}
\ \dot{\theta_{i}}=\omega_{i}+\sum\limits_{j=1}^N\lambda_{ij}A_{ij}\sin(\theta_{j}-\theta_{i}),
\quad i=1,2,\cdots,N,\
\end{equation}
where \(\theta_{i}\) denotes the phase of the \(i\)th oscillator, \(\omega_{i}\)s are the natural frequencies which are distributed by an even and symmetric probability density function  \(g(\omega)\). \(A_{ij}\) is the adjacency matrix of the network and \(\lambda_{ij}\) is the coupling strength between nodes \(i\) and \(j\), which are normally considered as \(\lambda_{ij}=\lambda/k_{i}\) , where \(\lambda\) is the global coupling strength and \(k_{i}\) is the degree of node \(i\) .

To quantify the synchronous behaviors of the oscillators, Kuramoto introduced the following order parameter~\cite{Kuramoto1984}

\begin{equation}\label{order}
\ R(t)e^{i\psi(t)}=\frac{1}{N}\sum\limits_{j=1}^Ne^{i\theta_{j}(t)}.
\end{equation}
The order parameter \(R(t)\) can be considered as the proportion of synchronized oscillators, and \(R(t)\in [0,1]\). When the coupled Kuramoto oscillators are synchronized, the order parameter will reach the value \(R=1\), while for completely incoherent situations, \(R=0\).

Here we focus on the influence of the spacial exponent \(\alpha\) and coupling strength \(\lambda\) on the emergence of global synchronization. Figure 2 shows how the dynamical parameter \(\lambda\) and the topological parameter \(\alpha\) affect the order parameter \(R\) . In the contour graph, the colour represents the value of order parameter \(R\) . The red region means the network reaches synchronization, and the blue one indicates that the phases of network oscillators are out of synchronization. Other colours denote that only partial nodes reach synchronization. As we can see, when the coupling strength \(\lambda\) is lager than some critical value \(\lambda_{c}\), phase synchronization appears at \(\alpha<\alpha_{op}\). In the contour graph, when \(\alpha>3\) , the network can hardly synchronize for any value of \(\lambda\) . As mentioned before, for Li networks, when \(\alpha=\alpha_{op}=d+1\) (for the considered lattice, \(d=2\) ), the small-world property emerges in the networks. So we can say that proper addition of long-range connections on a square lattice network, where the network topology is varying between the small-world type (\(\alpha=d+1\)) and the random one (\(\alpha=0\)), can greatly improve synchronizability of the network.

\begin{figure}[!ht]
\centering
\includegraphics[width=10cm]{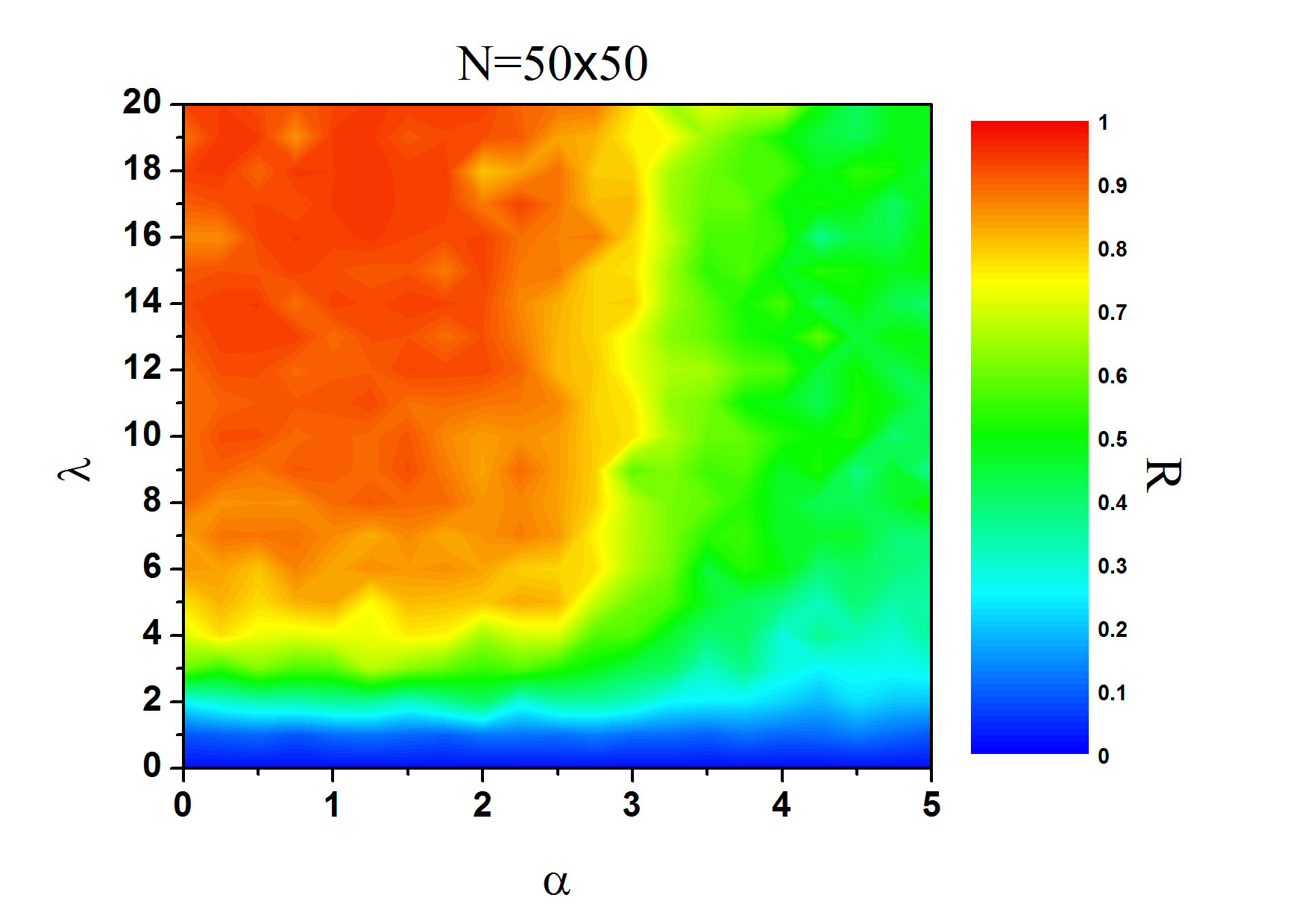}\\
\caption{The colour variation shows the value of order parameter \(R\) as a function of spacial exponent \(\alpha\) and coupling strength \(\lambda\), for a single layer Li networks, and the size of the network is \(N=50\times50\). For \(\alpha<3\) and \(\lambda>4\), phase synchronization can emerge. For \(\alpha>3\) or \(\lambda<4\), the network can hardly synchronize. }\label{single layer order}
\end{figure}

\subsection{\label{sec:level2}Synchronization on duplex Li network}
In order to investigate synchronous behaviors on multilayer spacial networks, we introduce a duplex Li network model. As shown in Fig. 3, the duplex network model has two layers and the nodes of each layer have the same geographic positions. The way intra-links connect intra-nodes follows the same rule as that for a single layer Li network. That is, the probability that two nodes locating in the same layer establish a long-range connection follows \(P(r_{ij})\sim r^{-\alpha}_{ij}\) , where \(r_{ij}\) is the Manhattan distance between nodes \(i\) and \(j\), and \(\alpha\) is the spacial exponent. In particular,  we set \(\alpha=d+1=3\) (the network exhibits small-world properties) in both layers. Moreover, the inter-links are one-to-one, that is, each node in one layer is connected to a counterpart in the other layer.

\begin{figure}[!ht]
  \centering
 \includegraphics[width=10cm]{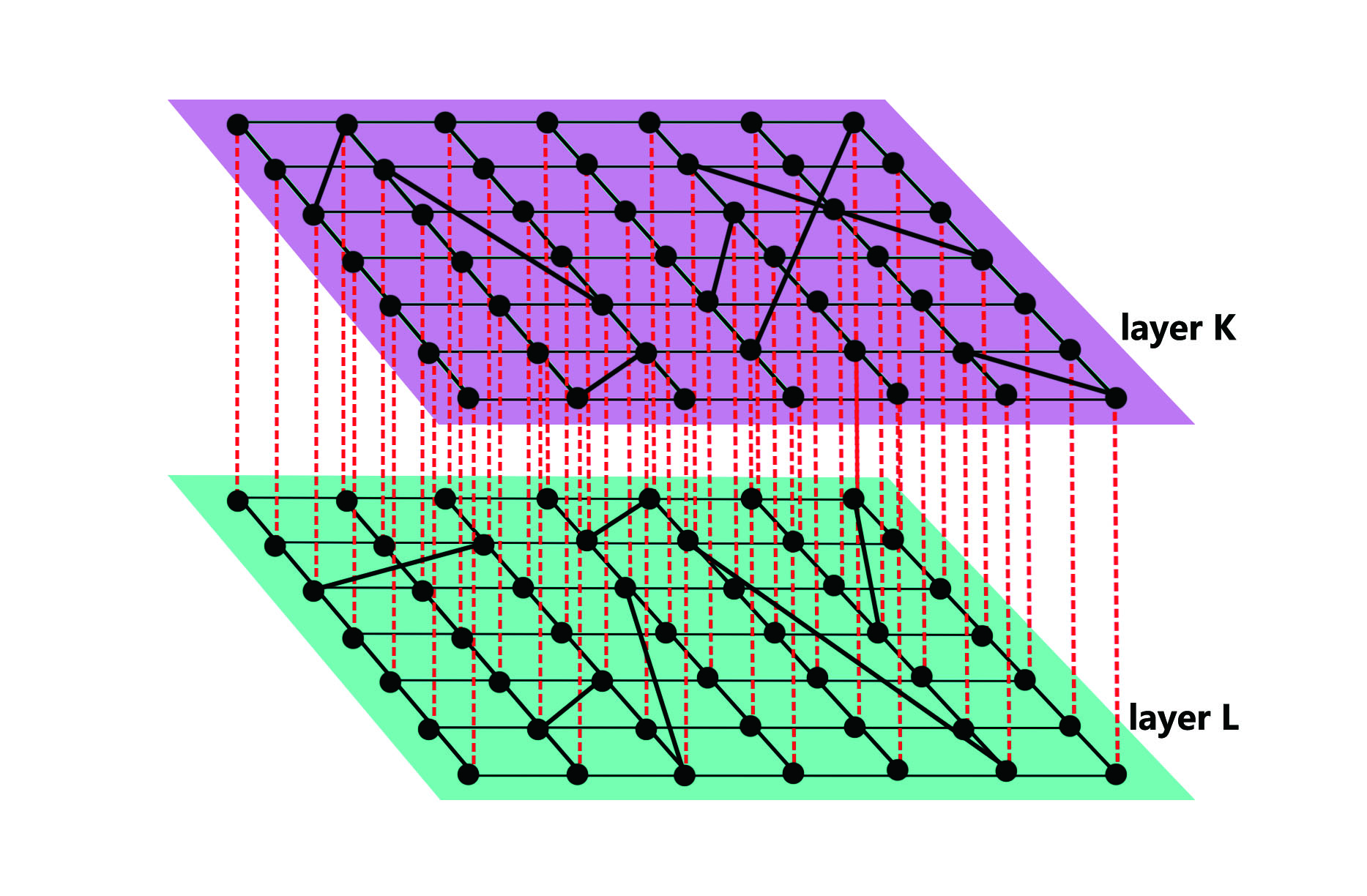}\\
     \caption{A duplex network consisting of layer \(K\) and layer \(L\). Each layer is a two-dimensional Li network, and each node in a layer has a connection with its counterpart in the other layer.}\label{duplex}
\end{figure}

Analogously, for the duplex Li network, the oscillators evolve by the Kuramoto model,

\begin{equation}\label{duplex kuramoto1}
\ \dot{\theta}_{i}^{K}=\omega_{i}^{K}+\frac{\lambda^{K}}{k_{i}}\sum\limits_{j\in\Lambda_{i}}\sin(\theta_{j}^{K}-\theta_{i}^{K})+\lambda^{KL}\sin(\theta_{i}^{L}-\theta_{i}^{K}),
\quad i=1,2,\cdots,N,\
\end{equation}
\begin{equation}\label{duplex kuramoto2}
\ \dot{\theta}_{i}^{L}=\omega_{i}^{L}+\frac{\lambda^{L}}{k_{i}}\sum\limits_{j\in\Lambda_{i}}\sin(\theta_{j}^{L}-\theta_{i}^{L})+\lambda^{KL}\sin(\theta_{i}^{K}-\theta_{i}^{L}),
\quad i=1,2,\cdots,N,\
\end{equation}
where \(\theta_{i}^{K}(\theta_{i}^{L})\) denotes the phase of the \(i\)th oscillator in layer \(K(L)\) , \(\omega_{i}^{K}(\omega_{i}^{L})\)s are the natural frequencies of layer \(K(L)\)'s oscillators being distributed by an even and symmetric probability density. \(\lambda^{K}(\lambda^{L})\) is the coupling strength of layer \(K(L)\) and \(\lambda^{KL}\) is the inter-layer coupling strength. \(\Lambda_{i}\) represents the set of neighbors of node \(i\) within the same layer. Using the global order parameter \(R_{global}\) , which is considered as the proportion of synchronized oscillators in both layers, we can find out whether two layers evolve into the same synchronous state. For \(R_{global}=1\), oscillators in both layers are fully synchronized, while for \(R_{global}=0\), most of the oscillators are not synchronized.

\begin{figure}[!ht]
  \centering
 \includegraphics[width=10cm]{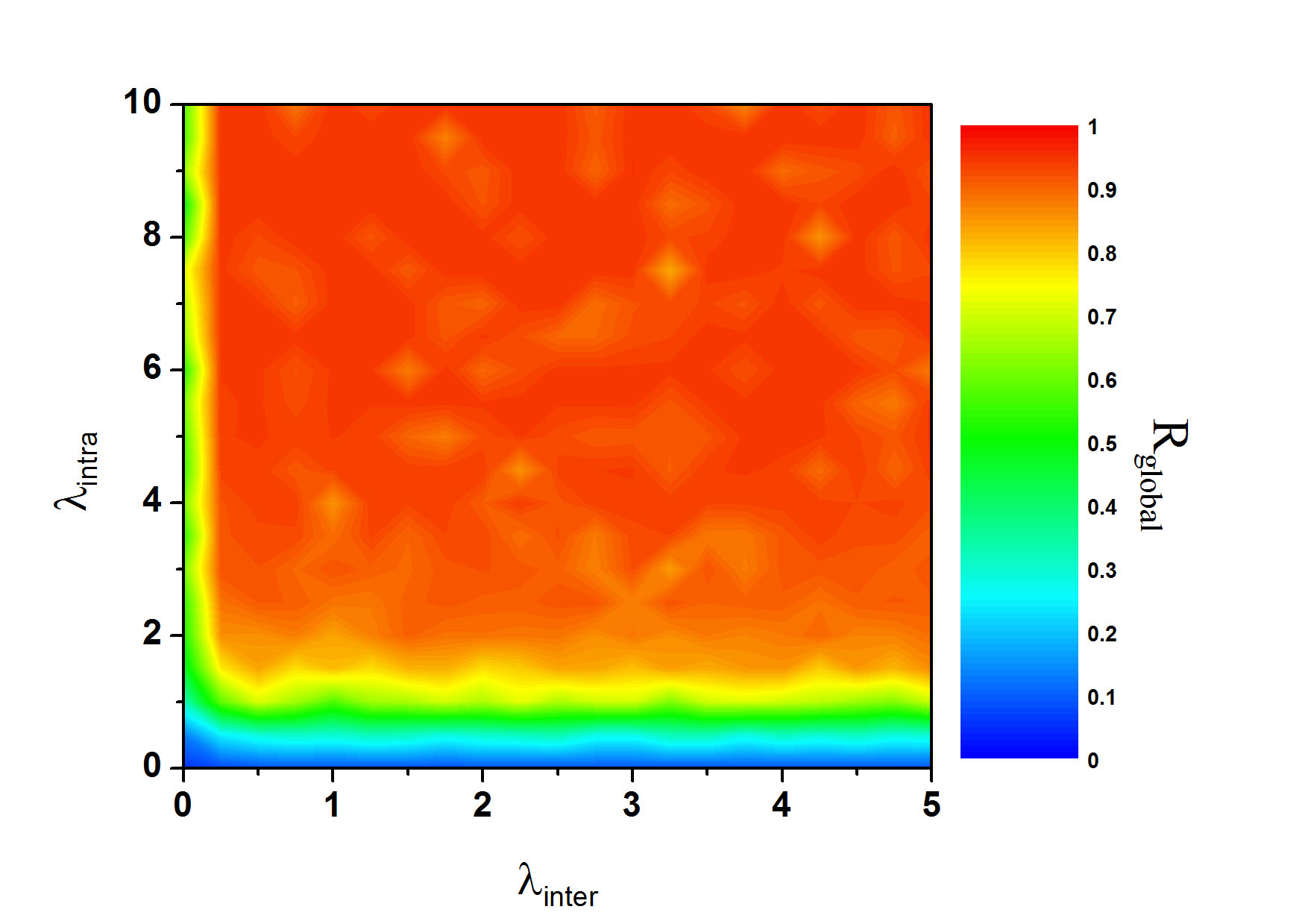}\\
     \caption{The color variation shows the value of global order parameter \(R_{global}\) as a function of inter-layer coupling strength \(\lambda_{inter}\) and intra-layer coupling strength \(\lambda_{intra}\), for a duplex Li network, where the size of each layer is \(N=10\times10\), and the spacial exponents of layer \(K\) and \(L\) are \(\alpha_{K}=\alpha_{L}=3\). In the region of \(\lambda_{inter}>0.25\) and \(\lambda_{intra}>2\), the network can evolve into global synchronization.}\label{duplex order}
\end{figure}

 For clarity, we set \(\lambda^K=\lambda^L=\lambda_{intra}\),  and \(\lambda^{KL}=\lambda_{inter}\). Figure 4 displays the influence of intra-layer coupling strength \(\lambda_{intra}\) and inter-layer coupling strength \(\lambda_{inter}\) on global synchronization, where \(\alpha=3\). We can see that the two factors have much impact on the order parameter \(R_{global}\). For \(\lambda_{inter}>0.25\) and \(\lambda_{intra}>2\), the duplex network's synchronizability is greatly improved. For other values of \(\alpha<3\), there exist similar results. In other words, large values of inter- and intra-coupling strength can lead duplex spacial networks into global synchronization.

\subsection{\label{sec:level2}Synchronization process on duplex Li network}
As is well known, for Kuramoto oscillators, full synchronization means the phases of all nodes modulo \(2\pi\) are the same. We can say that they all come to an identical state~\cite{Kuramoto1984}. While for multiplex networks, there are other kinds of synchronous states, as shown in Fig. 5. Intra-layer synchronization means all nodes within each layer have the same value of phase modulo \(2\pi\) , and inter-layer synchronization means each node in a layer reaches the same state as its counterparts in the other layer. In this section, we put forward a new order parameter to describe the inter-layer synchronization as follows,

\begin{equation}\label{inter}
\ R_{inter}=\frac{\sum\limits_{i=1}R(i)}{N}, R(i)>0.99.\
\end{equation}
Here, \(R(i)\) is the local order parameter between node \(i\) in one layer and its counterpart in the other layer, \(N\) is the total number of pairs of nodes between layers. We term \(R_{inter}\) as the inter-layer order parameter. For \(R_{inter}=1\) , the system reaches inter-layer synchronization, while for \(R_{inter}=0\) , it means no inter-layer synchronization appears.

\begin{figure}[!ht]
  \centering
 \includegraphics[width=16cm]{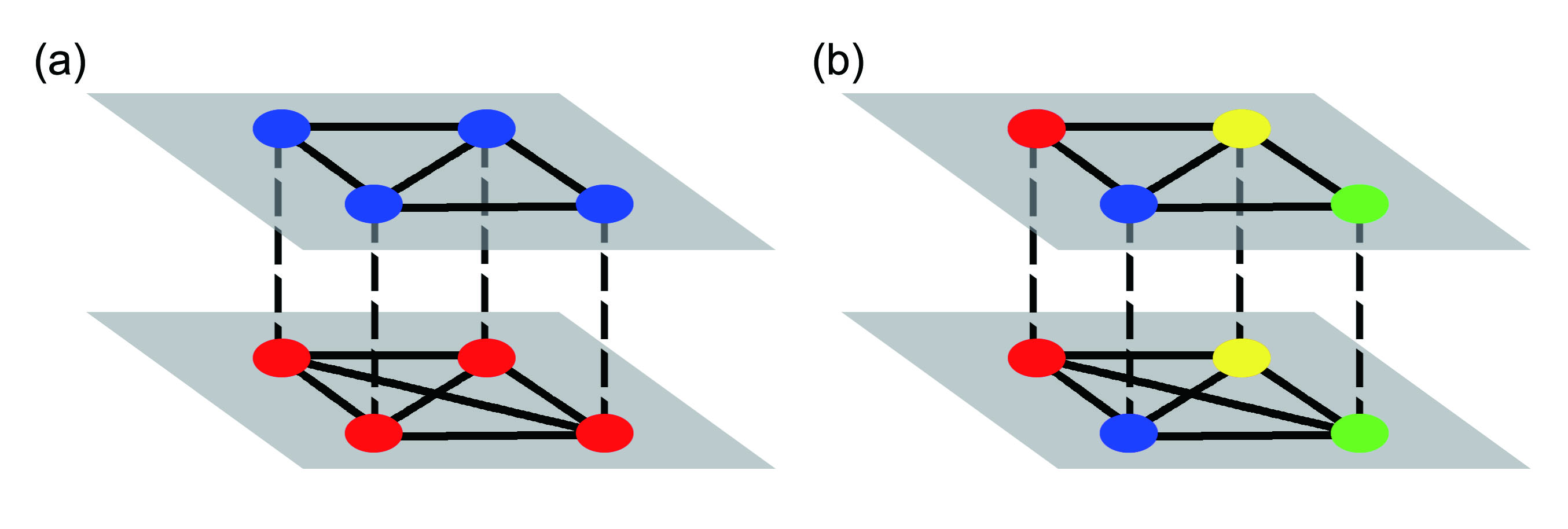}\\
     \caption{Schematic representation of (a) intra-layer synchronization and (b) inter-layer synchronization, in a duplex spacial network.}\label{duplex interorder}
\end{figure}

Figure 6 shows the impact of inter-layer and intra-layer coupling strength on the inter-layer order parameter. It is obvious that the inter-layer coupling strength \(\lambda_{inter}\) is a very important factor influencing inter-layer synchronization. When \(\lambda_{inter}\) is getting larger, the inter-layer order parameter comes closer to value \(1\). Also, larger intra-layer coupling strength \(\lambda_{intra}\) also lead to better inter-layer synchronizability.

\begin{figure}[!ht]
  \centering
 \includegraphics[width=10cm]{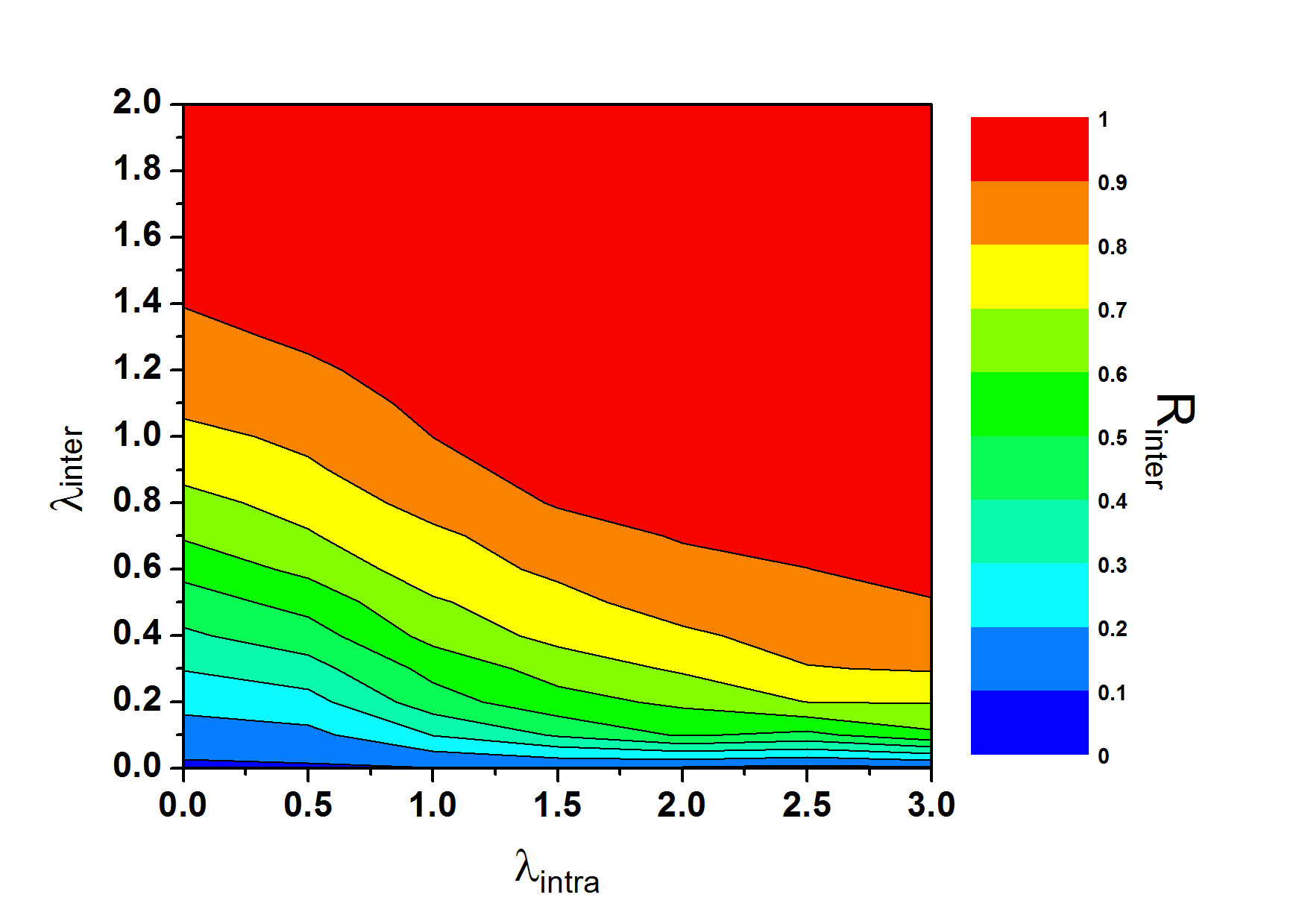}\\
     \caption{Inter-layer order parameter \(R_{inter}\) varying as a function of inter-layer coupling strength \(\lambda_{inter}\) and intra-layer coupling strength \(\lambda_{intra}\). For a duplex Li network, where the size of each layer is \(N=10\times10\), and the spacial exponents for both layers \(K\) and \(L\) are \(\alpha_{K}=\alpha_{L}=3\). }\label{duplex interorder}
\end{figure}

Figure 7 shows the inter-layer, intra-layer and global synchronization order parameter for different values of inter- and intra-layer coupling strength, for a duplex Li network with \(N=10\times10\) and \(\alpha_{K}=\alpha_{L}=3\). The　intra-layer synchronization order parameters are defined the same as that in Eq. (2) for nodes within each layer. For a large inter-layer coupling strength \(\lambda_{inter}=5\) , no matter what value of intra-layer coupling strength \(\lambda_{intra}\) is, inter-layer synchronization can be achieved and always be the earlier one compared with intra-layer or global synchronization. Since the spacial exponents \(\alpha\) are identical in both layers, the intra-layer and global synchronization can be reached simultaneously. While for a quite small value of inter-layer coupling strength, such as that shown in Panel (d) for \(\lambda_{inter}=0.01\) , inter-layer  synchronization cannot occur, neither can global synchronization. For a large intra-layer coupling strength, layer \(K\) and layer \(L\) can reach intra-layer synchronization. In general, for duplex spacial networks consisting of the Kuramoto oscillators, inter-layer coupling strength is a crucial factor determining inter-layer as well as global synchronization. Analogously, intra-layer coupling strength is a determining factor for intra-layer synchronization.

\begin{figure}[!ht]
\centering
\includegraphics[width=8cm]{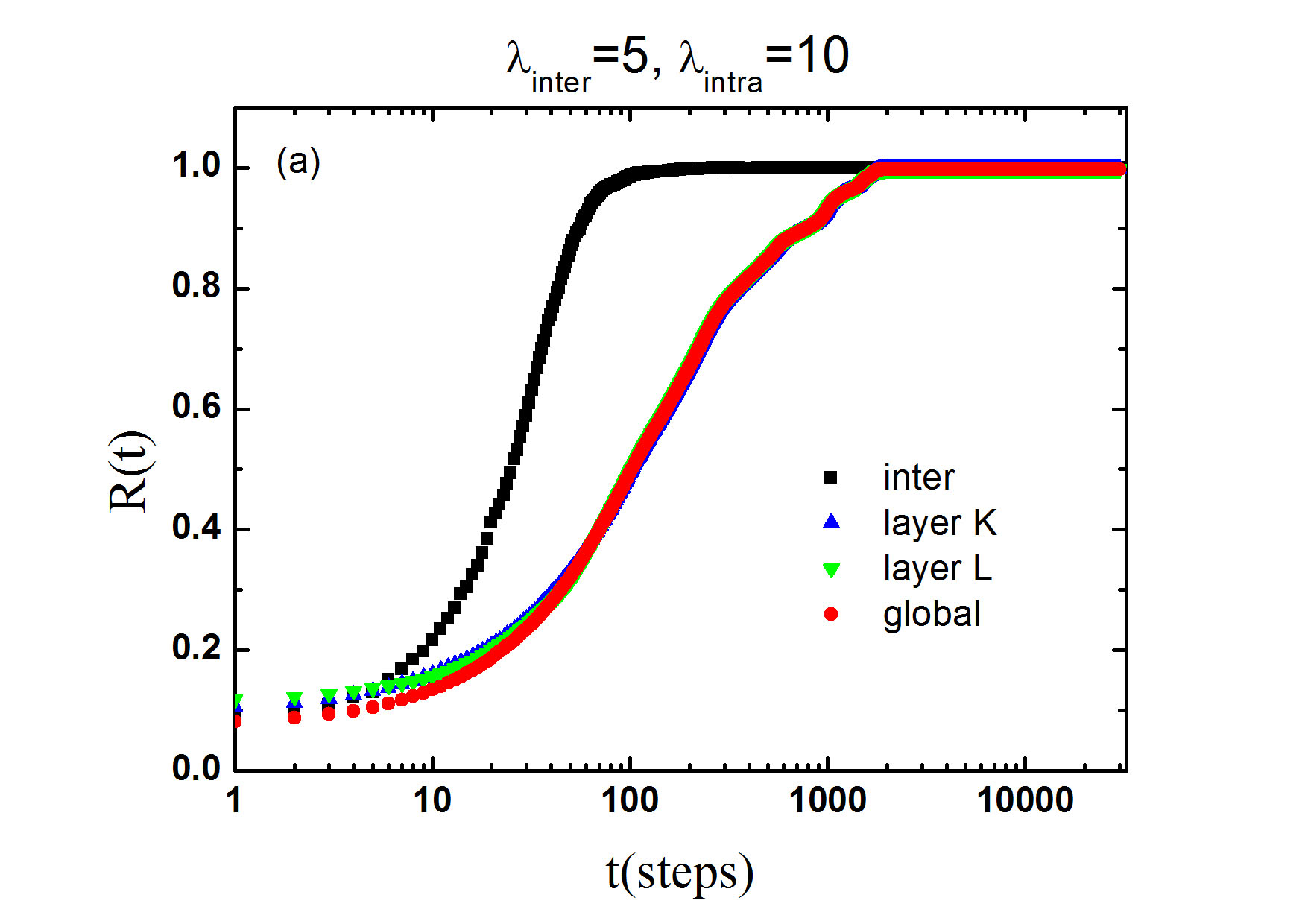}\includegraphics[width=8cm]{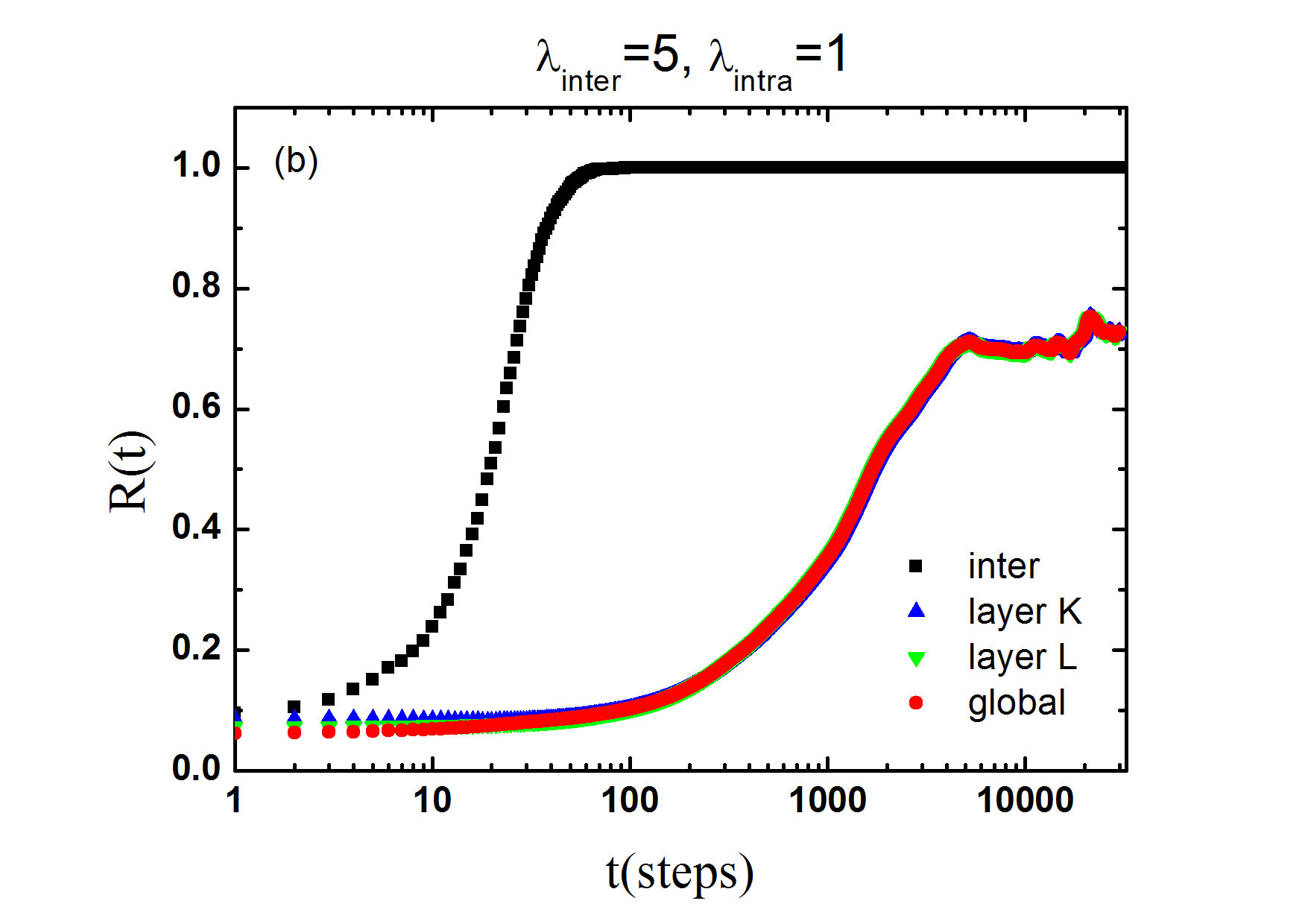}\\
\includegraphics[width=8cm]{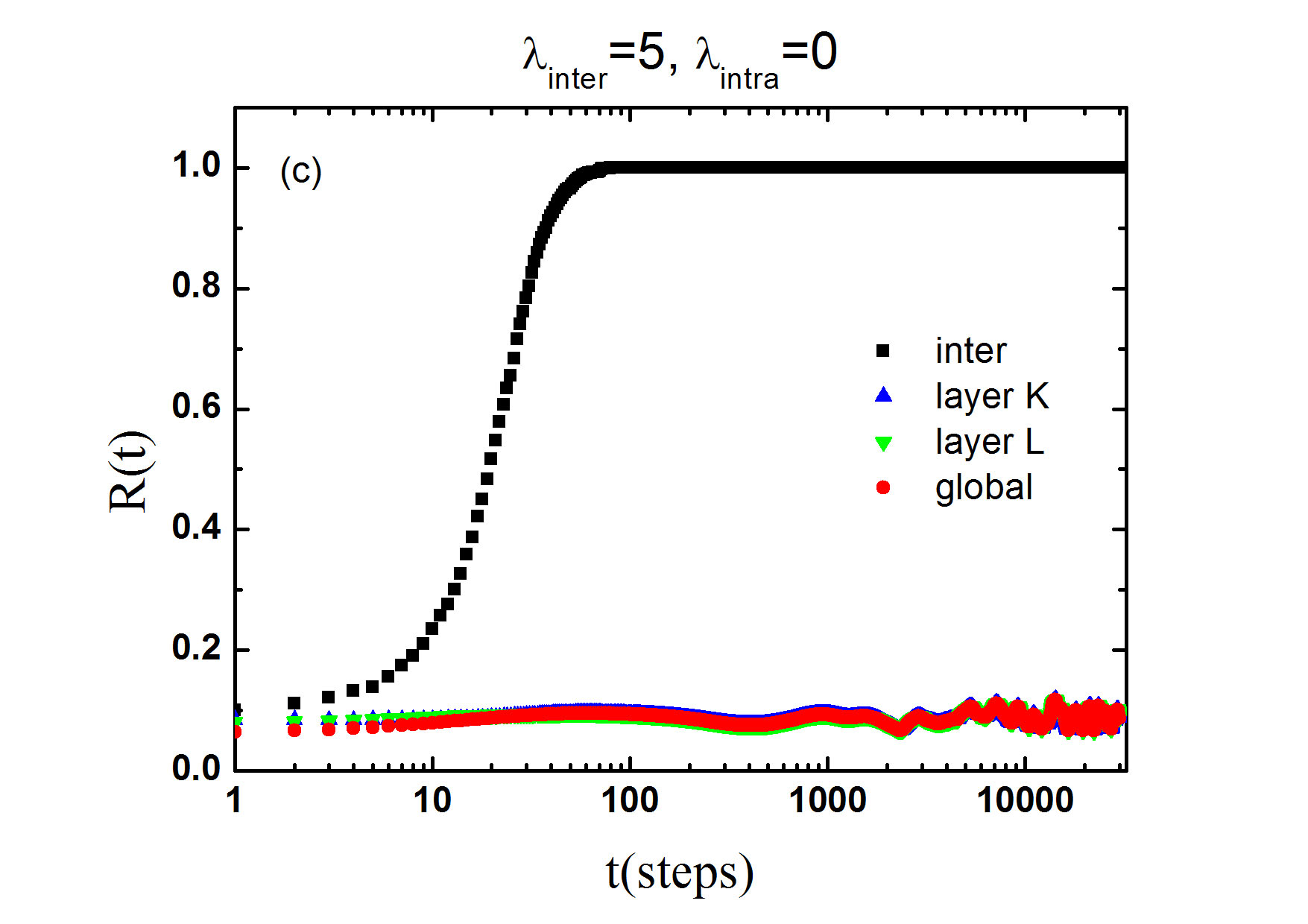}\includegraphics[width=8cm]{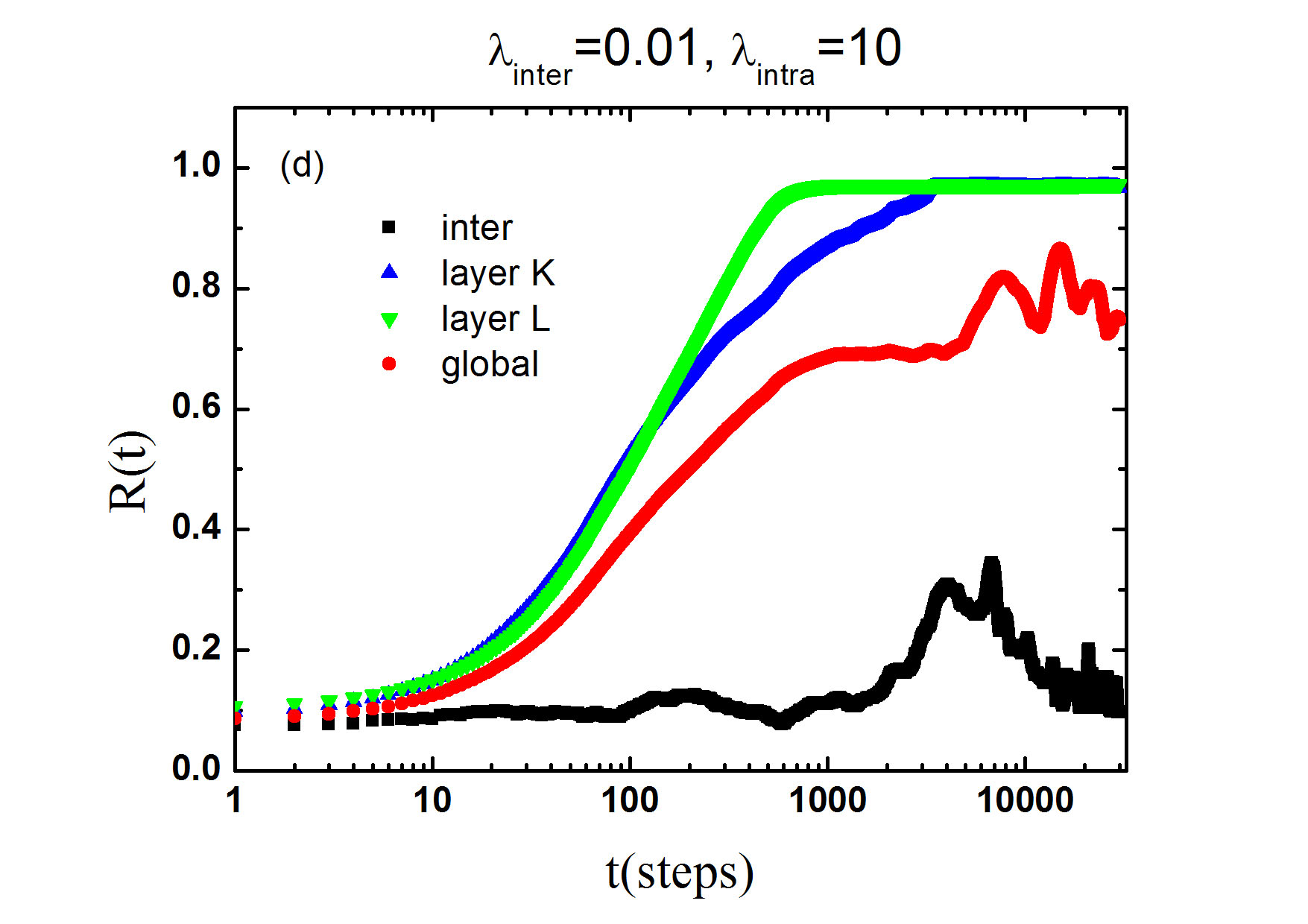}\\
\caption{The inter-layer, intra-layer and global order parameters varying with time \(t\). In each panel, the black squares represent the inter-layer order parameter, the blue upper triangles and green lower  triangles represent the intra-layer parameter of layer \(K\) and \(L\), respectively,  and the red circles represent the global order parameter. Panel (a): \(\lambda_{inter}=5\) and \(\lambda_{intra}=10\), the inter-layer order parameter reaches \(1\) when \(t\approx100\), while intra-layer and global order parameter reach \(1\) when \(t\approx1000\). Panel (b): \(\lambda_{inter}=5\) and \(\lambda_{intra}=1\), the inter-layer order parameter reaches \(1\) after \(t\approx100\), while intra-layer and the global order parameter is oscillating between 0.6 and 0.8, meaning that the duplex network reaches the lock phase state. Panel (c): \(\lambda_{inter}=5\) and \(\lambda_{intra}=0\), the inter-layer order parameter reaches \(1\) after \(t\approx100\), while intra-layer or global synchronization cannot be reached. Panel (d): \(\lambda_{inter}=0.01\) and \(\lambda_{intra}=10\), layer \(K\) and layer \(L\) can reach intra-layer synchronization, while neither inter-layer nor global synchronization happens.}\label{process}
\end{figure}

In order to further investigate how inter-layer coupling enhances the inter-layer synchronization, we illustrate the evolution progress of phase oscillators in Figs. 8 - 11. In all the figures, the horizontal axis denotes the evolution time steps and vertical axis represents the serial number of nodes in each layer. While the colours mean the value of phase modulo \(2\pi\) , and the same colour means the same synchronous state.  Here, for clearer and more visible observations, we use the round of phases modulo \(2\pi\) instead of phases modulo \(2\pi\).

For \(\lambda_{inter}=5\) and \(\lambda_{intra}=10\) shown in Fig. 8, all nodes in both layers reach synchronization after a certain time steps. Moreover, for clarity, we enlarge Panel (a) in the time slot [0, 2000], and plot the node dynamics in the way symmetric about \(t=0\), as shown in Panel (b). Therefore, the left half of this panel is for dynamical evolution of  nodes  in Layer \(K\), and  the right half is for that in Layer \(L\). It is obvious that,  due to the strong inter-layer interaction, each node in one layer is influenced by its counterpart in the other layer and reaches a synchronous state before its intra-layer neighbors do.

\begin{figure}[!ht]
\centering
\includegraphics[width=8cm]{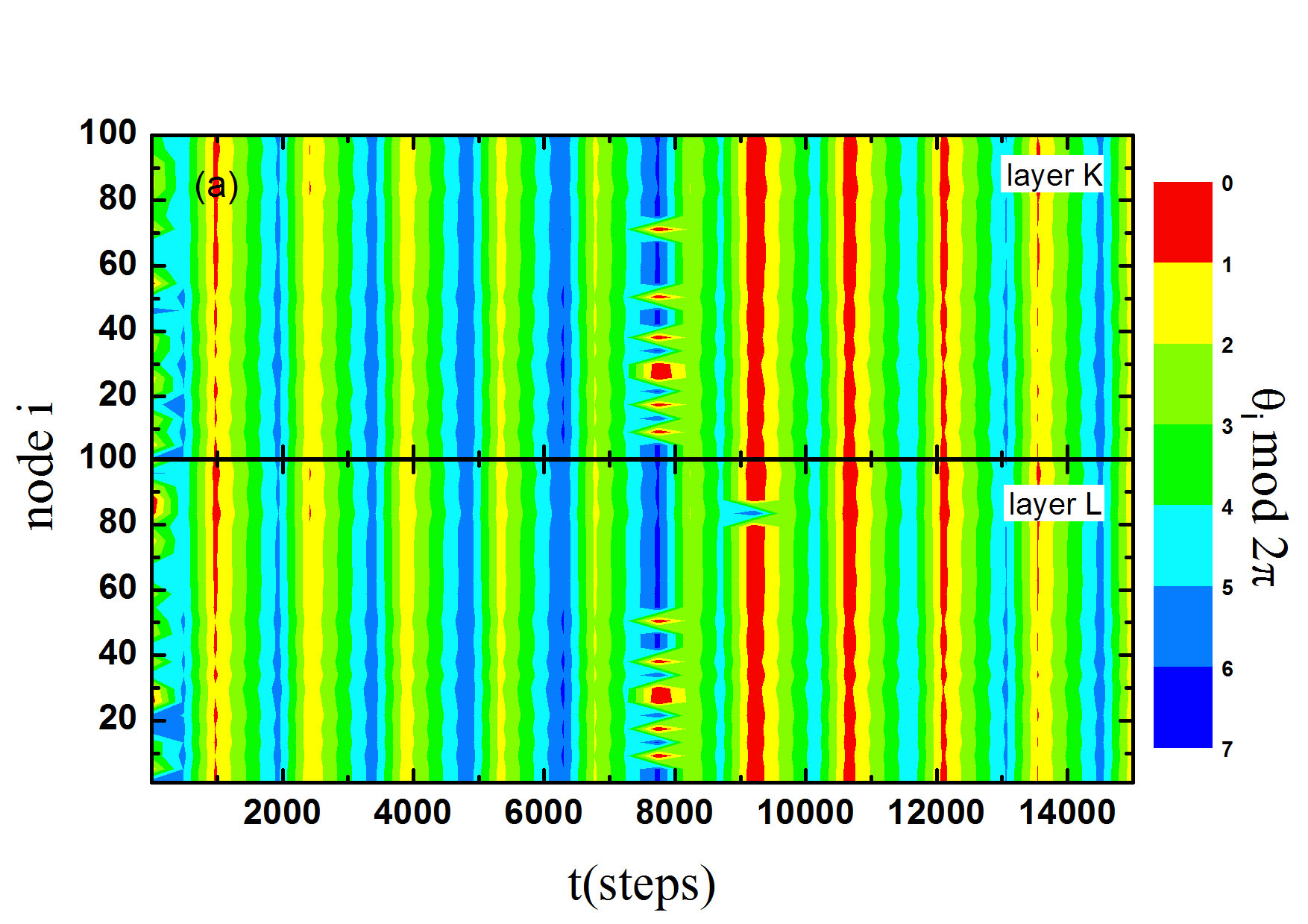}\includegraphics[width=8cm]{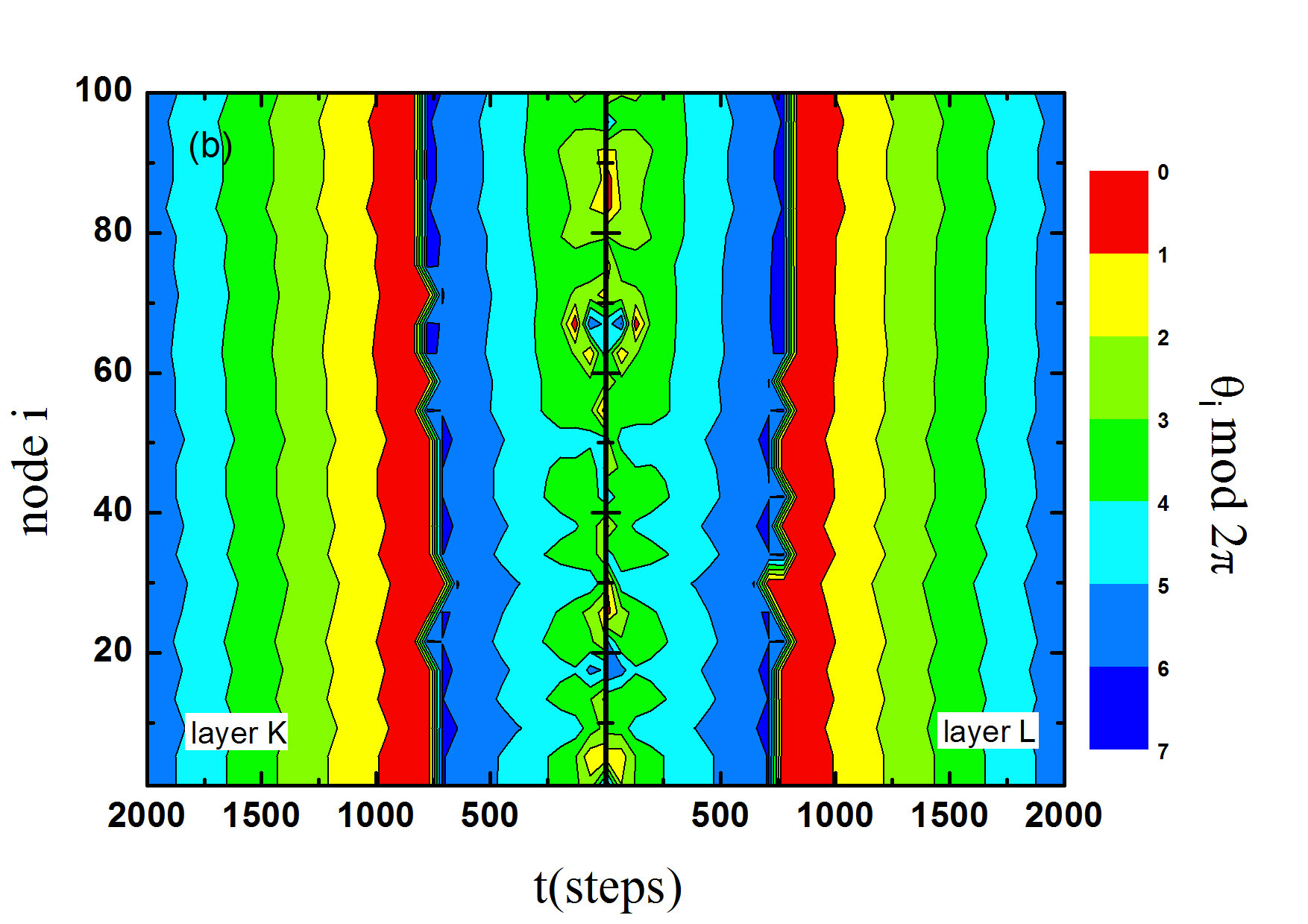}\\
\caption{Node dynamics in each layer evolving with time, for \(\lambda_{inter}=5\) and \(\lambda_{intra}=10\). (a): Each pair of nodes in different layers will arrive at the same state after \(t\approx1000\); (b): Enlarged plot in \(t\in[0,2000]\) of panel (a), each node and its counterpart in the other layer will get into the same state after a very short time period.}\label{theta}
\end{figure}

For \(\lambda_{inter}=5\) , \(\lambda_{intra}=1\) , Fig. 9 shows that, because of the weak intra-layer coupling strength, only partial synchronization occurs. We can obtain from the left panel that part of nodes in the two layers arrive at cluster synchronization, as  indicated  by the same colors emerging simultaneously. Similarly, in the zoom in Panel (b) of Fig. 9, each node in one layer and its counterpart in the other always stay in an identical state after some time steps.

\begin{figure}[!ht]
\centering
\includegraphics[width=8cm]{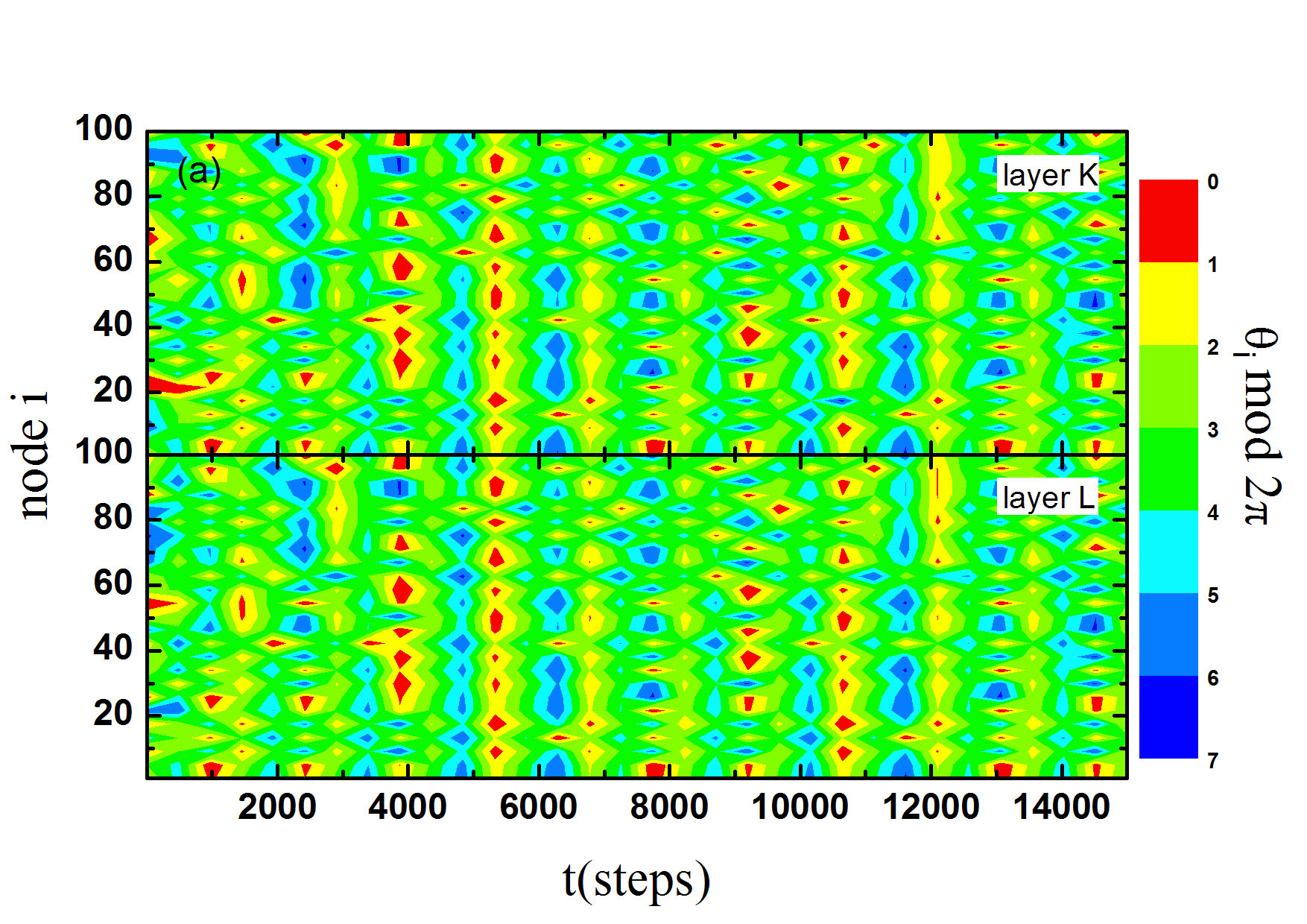}\includegraphics[width=8cm]{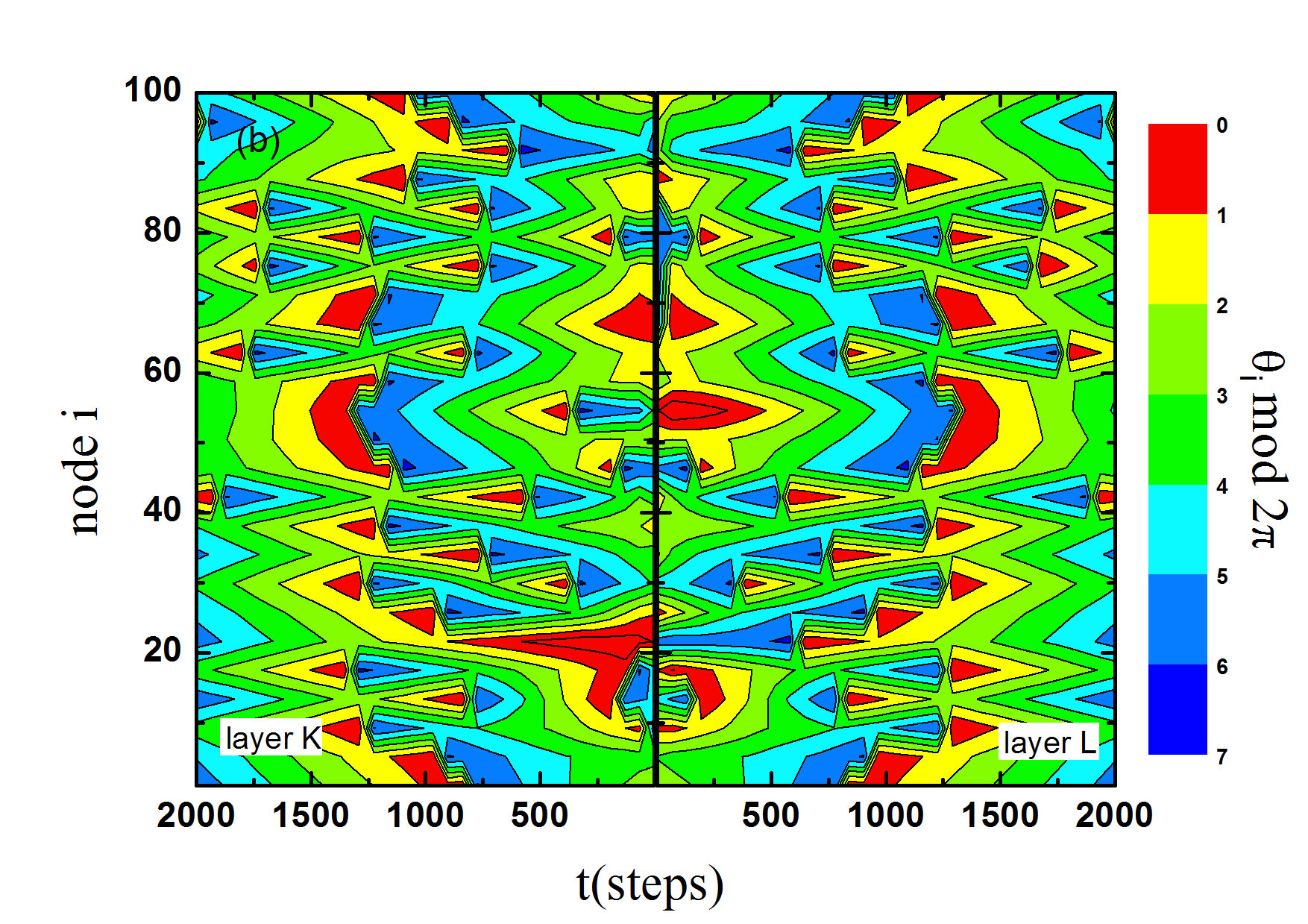}\\
\caption{Node dynamics in each layer evolving with time, for \(\lambda_{inter}=5\) and \(\lambda_{intra}=1\). (a): Only part of nodes will arrive at the same state at the same time; (b): Enlarged plot in \(t\in[0,2000]\) of (a), each node and its counterpart in the other layer will get into the same state after a very short time period.}\label{theta}
\end{figure}

When we set \(\lambda_{inter}=5\), and \(\lambda_{intra}=0\), we can observe that there is no occurrence of intra-layer or global synchronization, as shown in Fig. 10. From Panel (b), it is obvious that only inter-layer synchronization is achieved.

\begin{figure}[!ht]
\centering
\includegraphics[width=8cm]{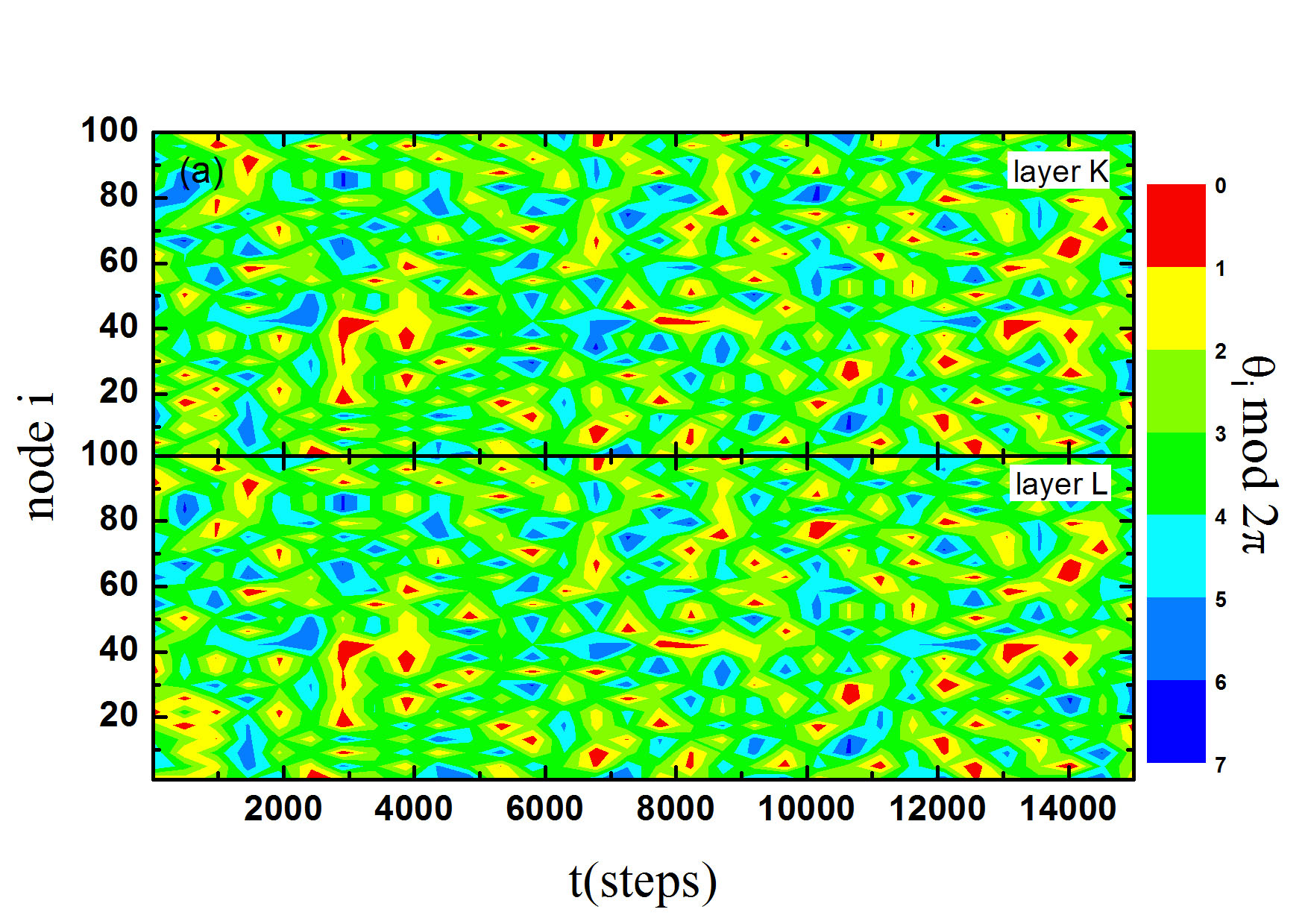}\includegraphics[width=8cm]{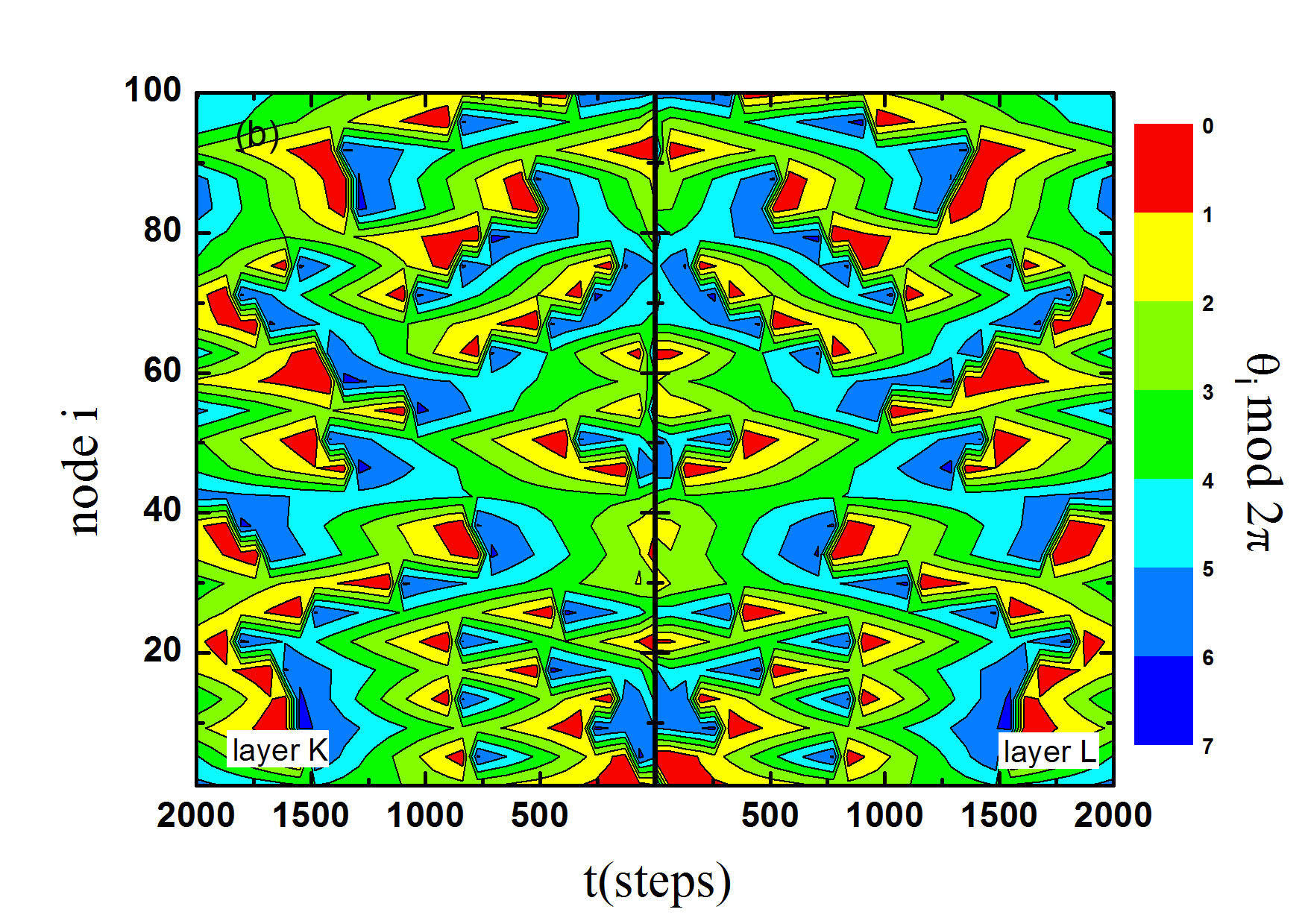}\\
\caption{Node dynamics in each layer evolving with time, for \(\lambda_{inter}=5\) and \(\lambda_{intra}=0\). (a): The nodes in the same layer or the other layer can not get into large synchronous cluster at the same time; (b): Enlarged plot in \(t\in[0,2000]\) of (a), each node and its counterpart in other layer will get into same state after a very short time period.}\label{theta}
\end{figure}

Figure 11 shows the case for strong intra-layer coupling but weak inter-layer coupling, with \(\lambda_{intra}=10\) and \(\lambda_{inter}=0.01\) . It is obvious that nodes in each layer reach its specific synchronous state, but no inter-layer synchronization occurs. From Panel (b) of Fig. 11, we can obtain that because of weak interaction between layers, a node in one layer can not reach the same state as its counterpart in the other layer. Figures 10 and 11 imply that both inter- and intra-layer interactions are crucial factors determining global synchronization.

\begin{figure}[!ht]
\centering
\includegraphics[width=8cm]{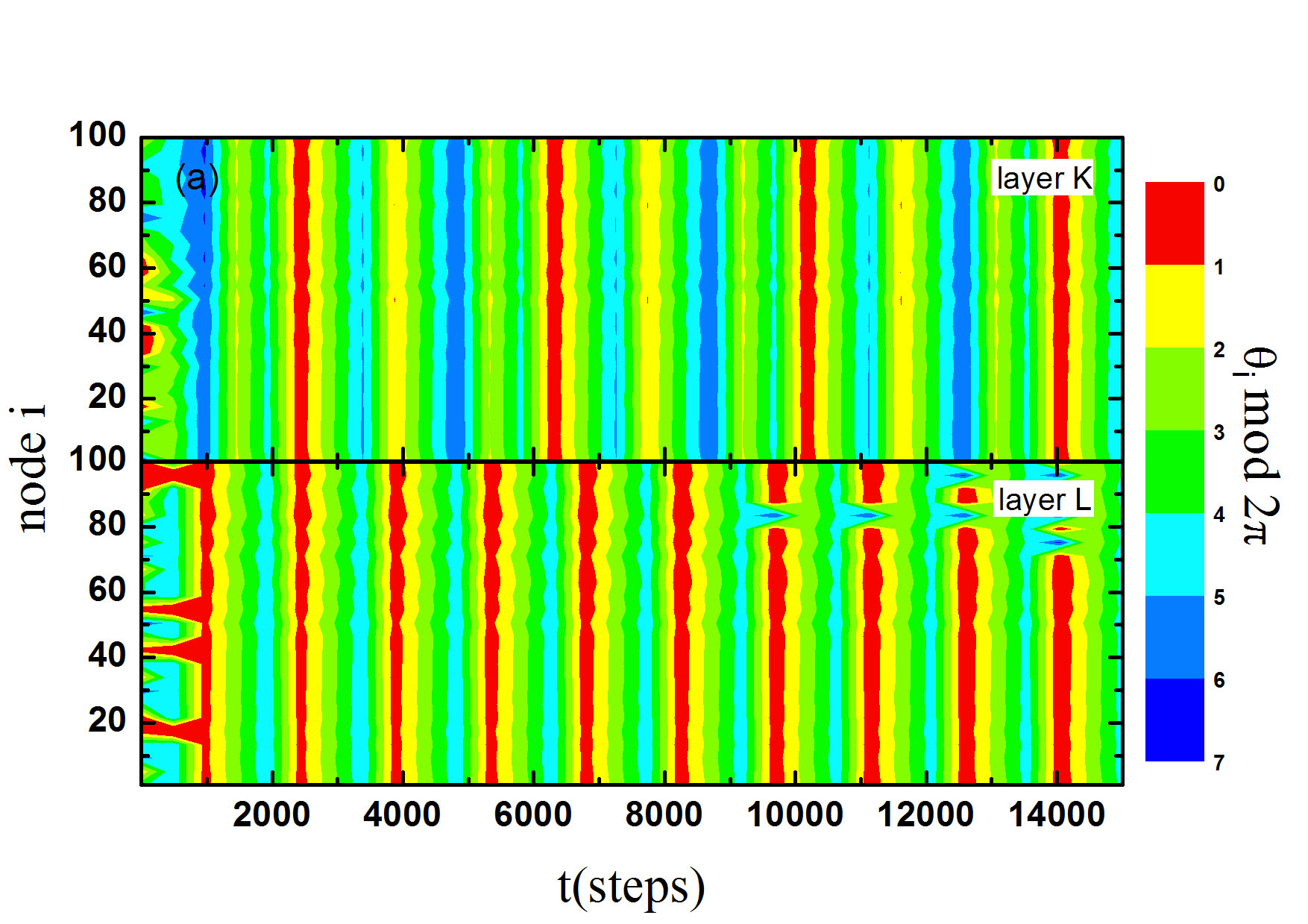}\includegraphics[width=8cm]{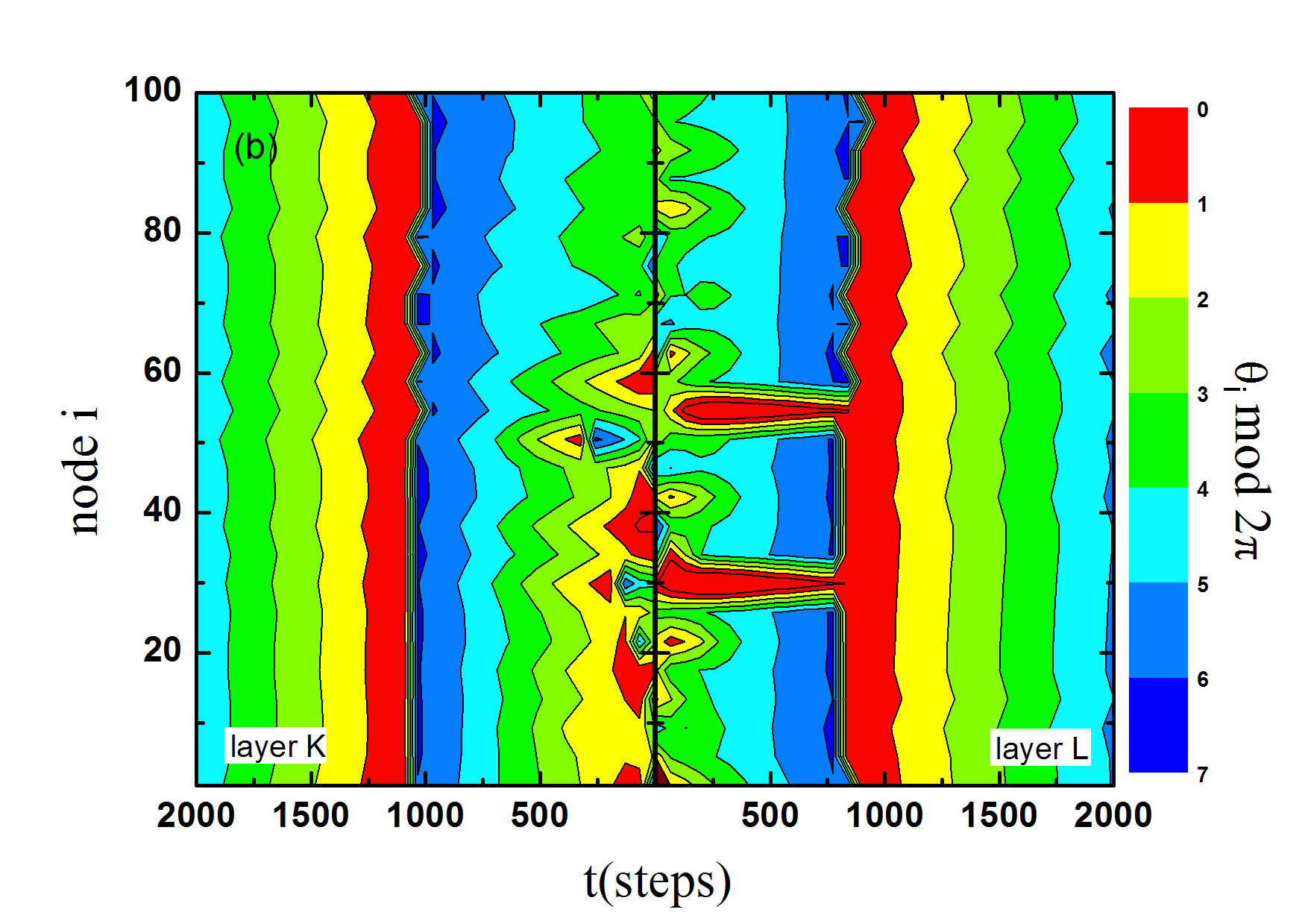}\\
\caption{Node dynamics in each layer evolving with time, for \(\lambda_{inter}=0.01\) and \(\lambda_{intra}=10\) .(a): The nodes from the same layer are in the same synchronous cluster, and the nodes from another layer are in different cluster; (b): Enlarged plot in \(t\in[0,2000]\) of (a), each node and its counterpart in the other layer can not get into same state at the same time.}\label{theta}
\end{figure}

Many results have indicated that, for scale-free networks, nodes with larger degrees are more likely to get into the synchronized state~\cite{GG2007}. Our study on single layer spacial small-world networks have the same results (Li networks with the spacial exponent being 3) yields similar observation. Panel (a) of Fig. 12 shows the likelihood for nodes of degree $k$ belonging to the largest synchronized connected component (GC) with respect to the coupling strength \(\lambda\) for single layer Li networks. The probability \(P_{GC}\) is defined by \(P_{GC}=n/m\), where \(m\) is the total round of simulation and \(n\) is the number of times when a node gets into the largest synchronized connected component. The panel indicates that highly connected nodes are more likely to reach synchronization state. While for duplex spacial networks, as shown in Panel (b), this phenomenon becomes much less obvious, which is possibly due to interaction between layers. In other words, interaction between layers can greatly affect the synchronization process.

\begin{figure}[!ht]
\centering
\includegraphics[width=8cm]{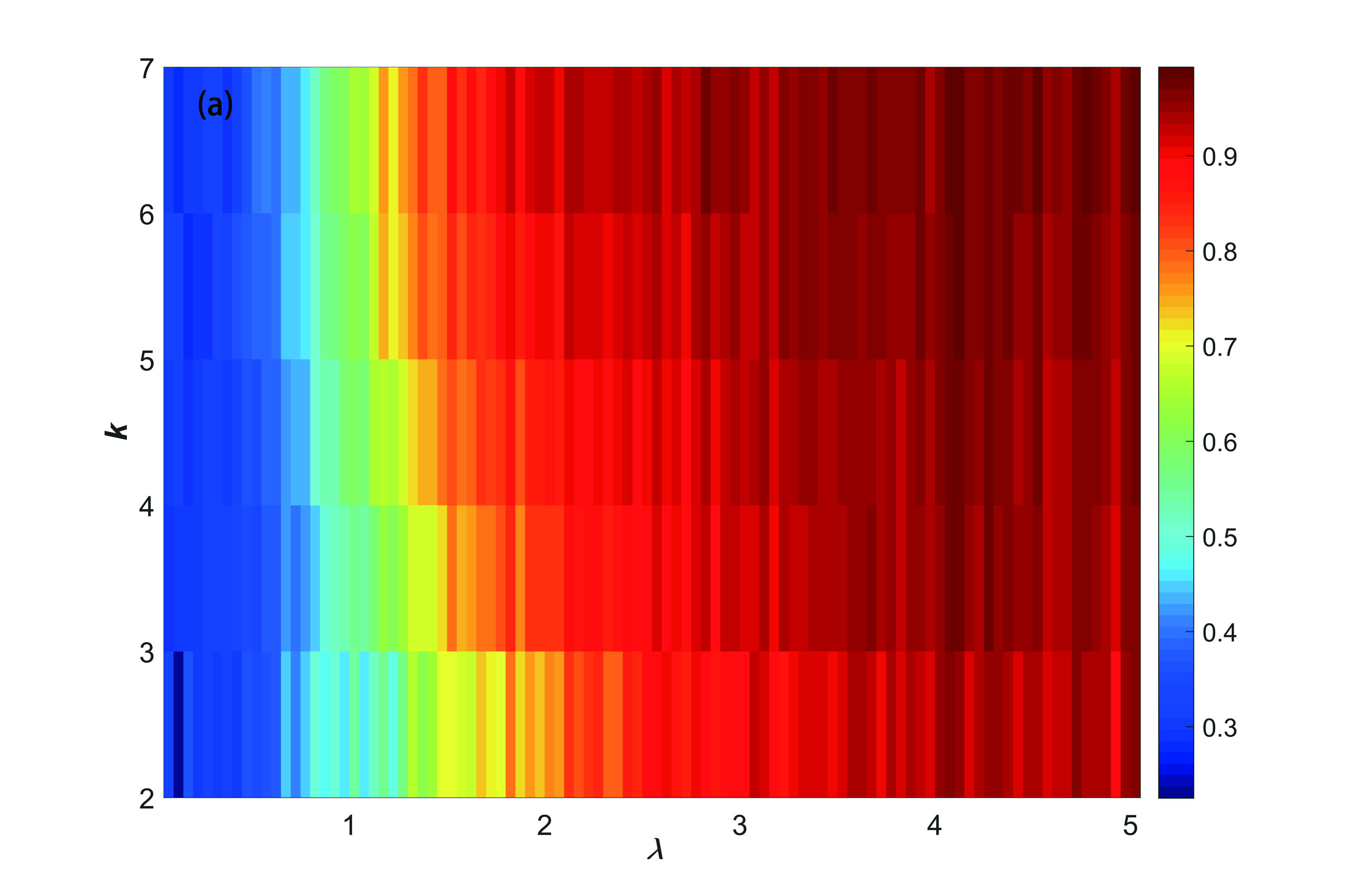}\includegraphics[width=8cm]{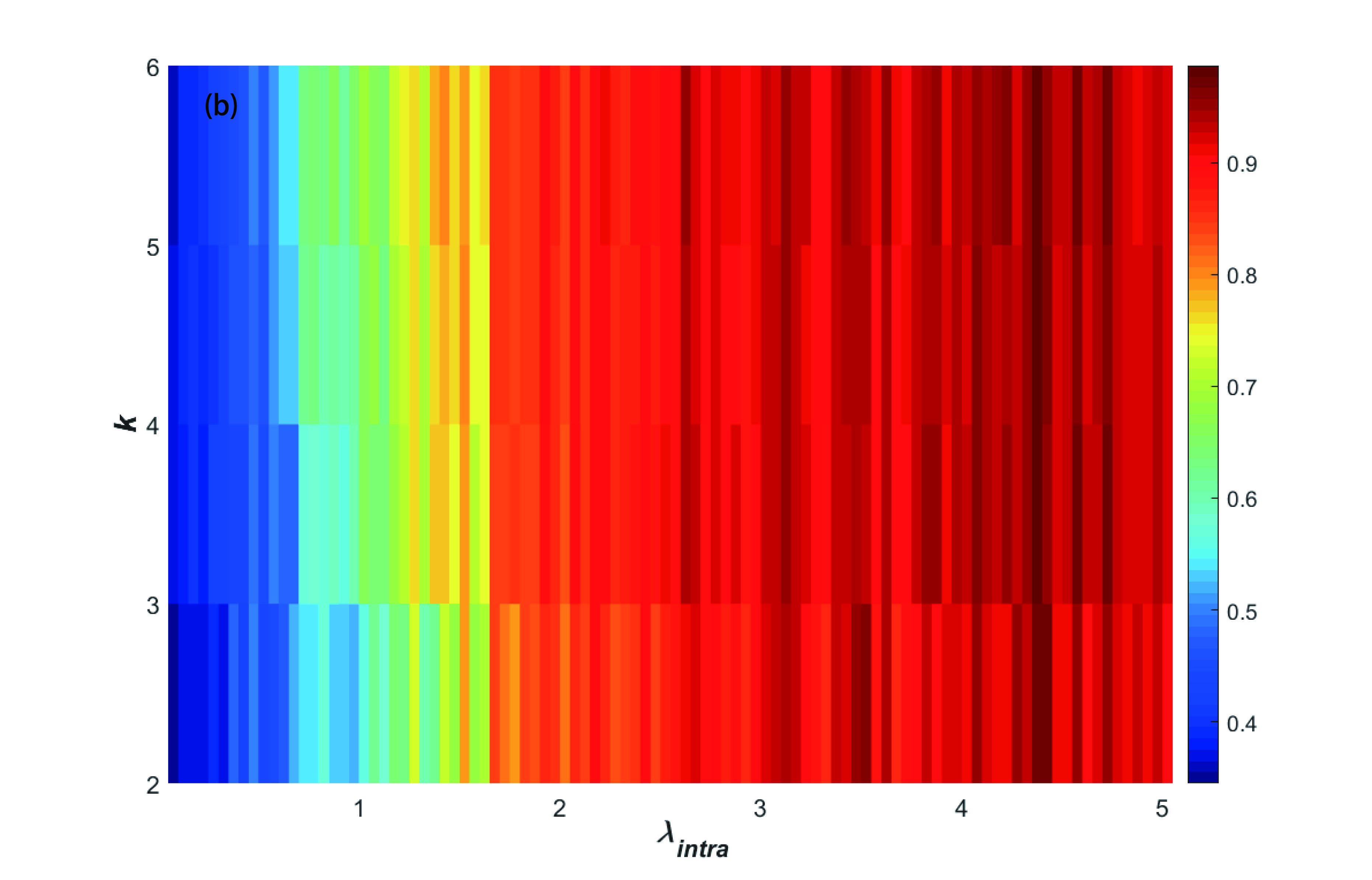}\\
\caption{The probability \(P_{GC}\) for a node of degree $k$ belonging to the largest synchronized connected component as a function of coupling strength \(\lambda\) (\(\lambda_{intra}\)) and \(\lambda_{inter}=10\). (a): In single layer Li networks, nodes with larger degrees will arrive at synchronization earlier; (b): For duplex Li networks, the influence of degrees will be weakened. }\label{degree}
\end{figure}

\subsection{\label{sec:level2}The impact of inter-links on synchronization}
As studied previously, the coupling strength between layers plays a dominant role in the emergence of global synchronization on multiplex networks. In other words, the interaction across layers can greatly change the synchronous behaviors of multiplex networks. In this section, we study how the density of links between layers (inter-links) influences the emergence of global synchronization. Fist of all, we use the previous duplex Li network model with one-to-one inter-links and randomly remove a fraction of inter-links by \(1-P_{link}\) , thus the density of inter-links can be denoted as \(P_{link}=M/N\) , where \(M\) is the number of links between layers after the removal and \(N\) is the size of each layer. Then we investigate the global order parameter with different inter-link density \(P_{link}\) in the parameter space of inter-  and intra-layer coupling strength, with results being shown in Fig. 13. As we can see, for a relatively small value of inter-link density such as \(P_{link}=0.1\), the synchronizability is just like the network before the removal (\(P_{link}=1\)).

\begin{figure}[!ht]
\centering
\includegraphics[width=8cm]{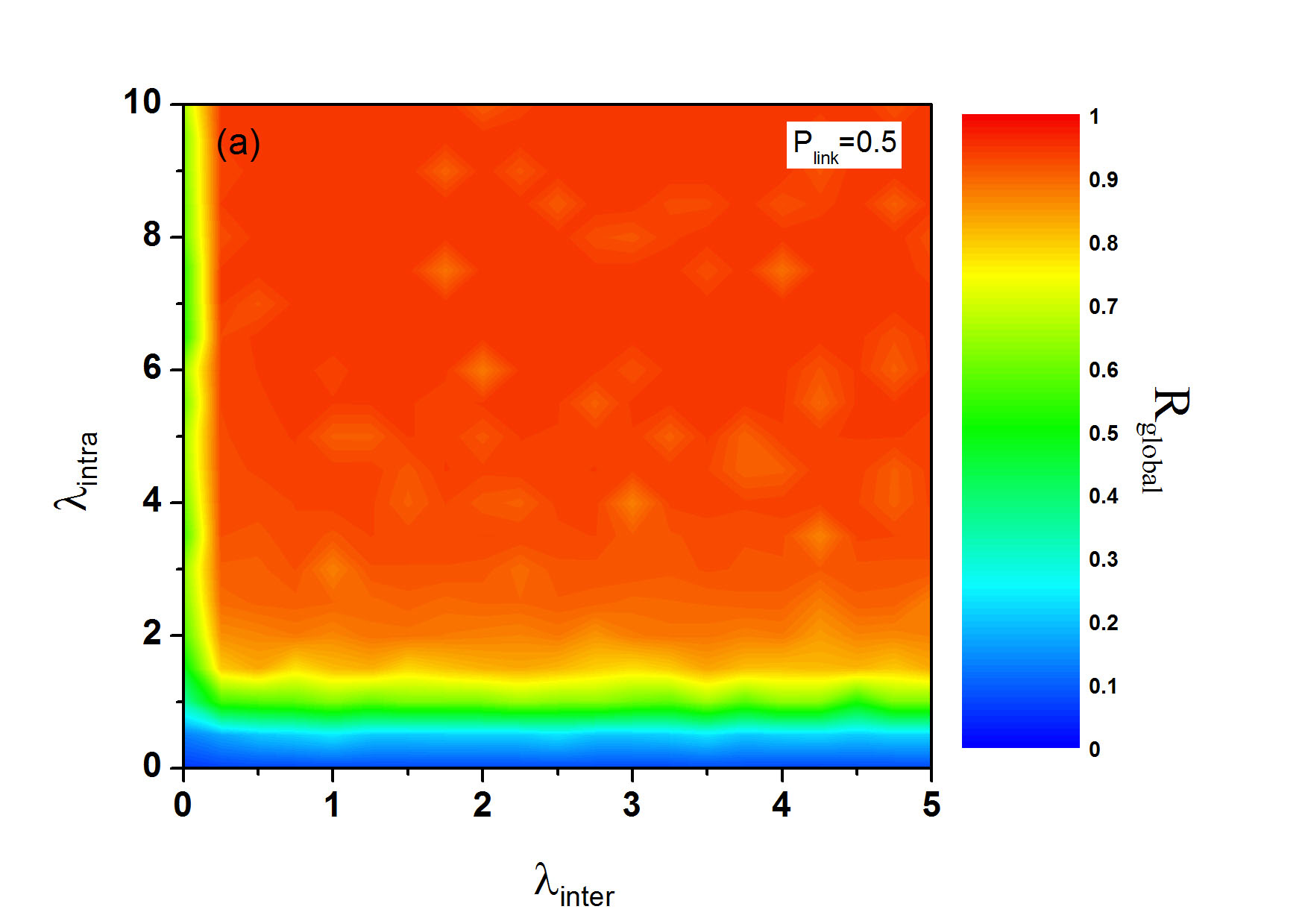}\includegraphics[width=8cm]{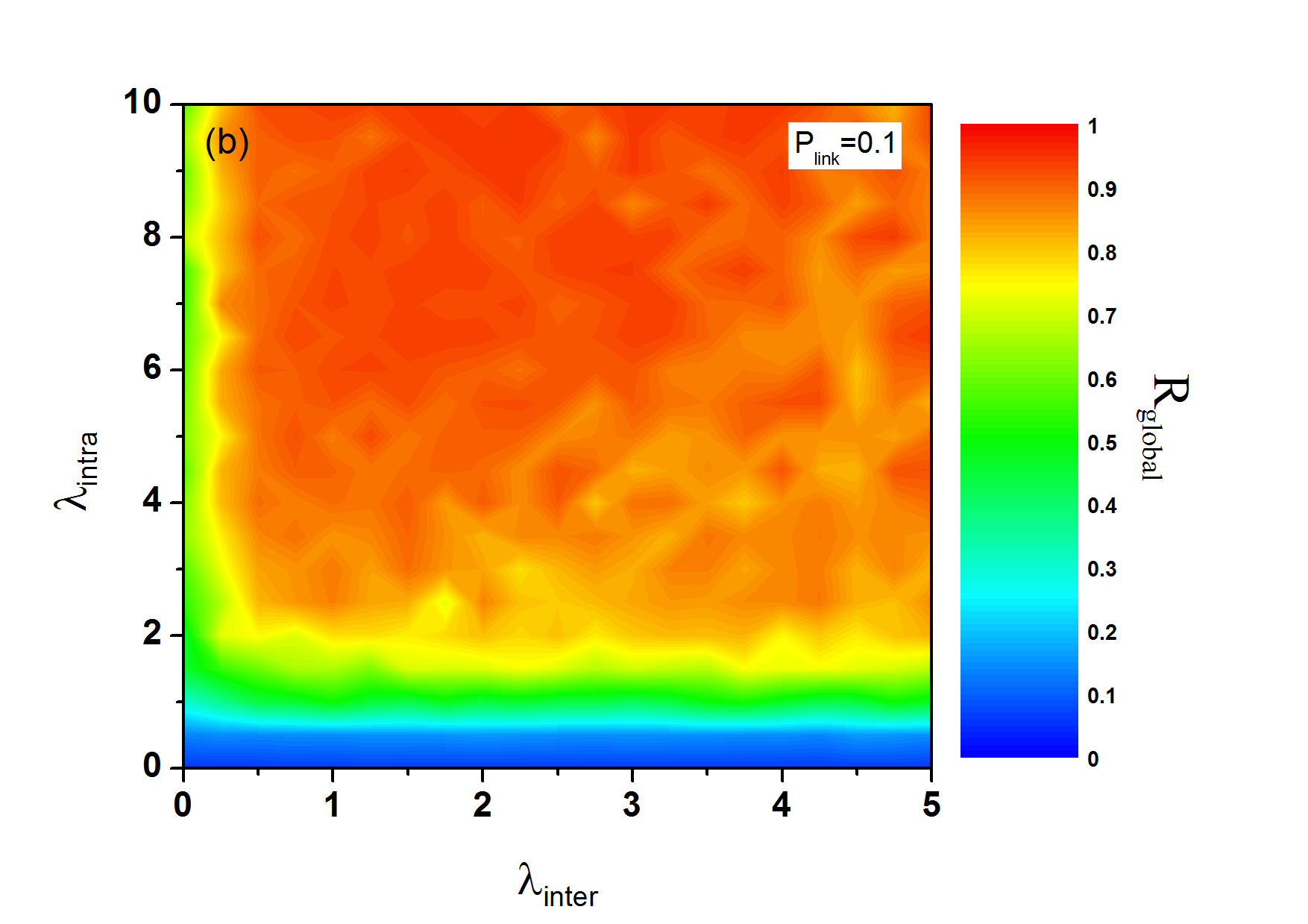}\\
\includegraphics[width=8cm]{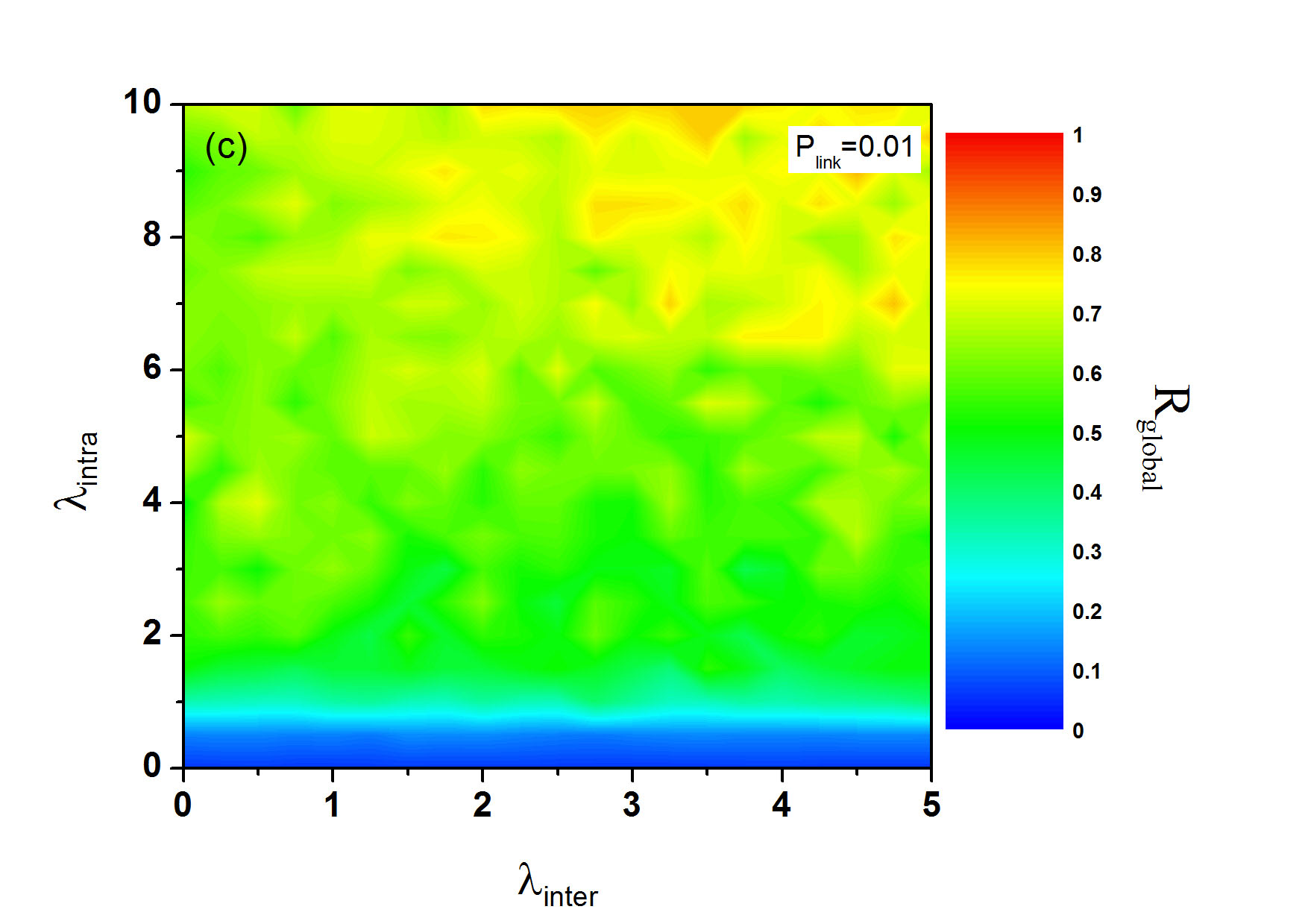}\\
\caption{Global order parameter \(R_{global}\) as a function of inter-layer coupling strength \(\lambda_{inter}\) and intra-layer coupling strength \(\lambda_{intra}\) for different inter-link densities \(P_{link}\). (a): \(P_{link}=0.5\), the synchronization region regarding \(\lambda_{inter}\) and \(\lambda_{intra}\) is similar to Fig. 4; (b): \(P_{link}=0.1\), the synchronization region begin to shrink; Panel (c). \(P_{link}=0.01\), there is no synchronization in the parameter space for $\lambda_{inter} \in [0,5]$ and  $\lambda_{intra} \in [0,10]$ .}\label{interlinks}
\end{figure}

In Fig. 14, we further illustrate different order parameters varying with inter-link density \(P_{link}\) , where \(\lambda_{inter}=\lambda_{intra}=5\). It is obvious that,  when we remove the inter-links between layers, there exists a threshold \(P_{link}^{c}\approx0.1\). For \(P_{link}>P_{link}^{c}\) , the network can always have inter-layer, intra-layer and global synchronization, while for \(P_{link}<P_{link}^{c}\) , only intra-layer synchronization appears. Therefore in duplex Li networks, a small value of inter-link density (as small as about 0.1 for this case ) can make the whole network reach global synchronization.

\begin{figure}[!ht]
  \centering
 \includegraphics[width=10cm]{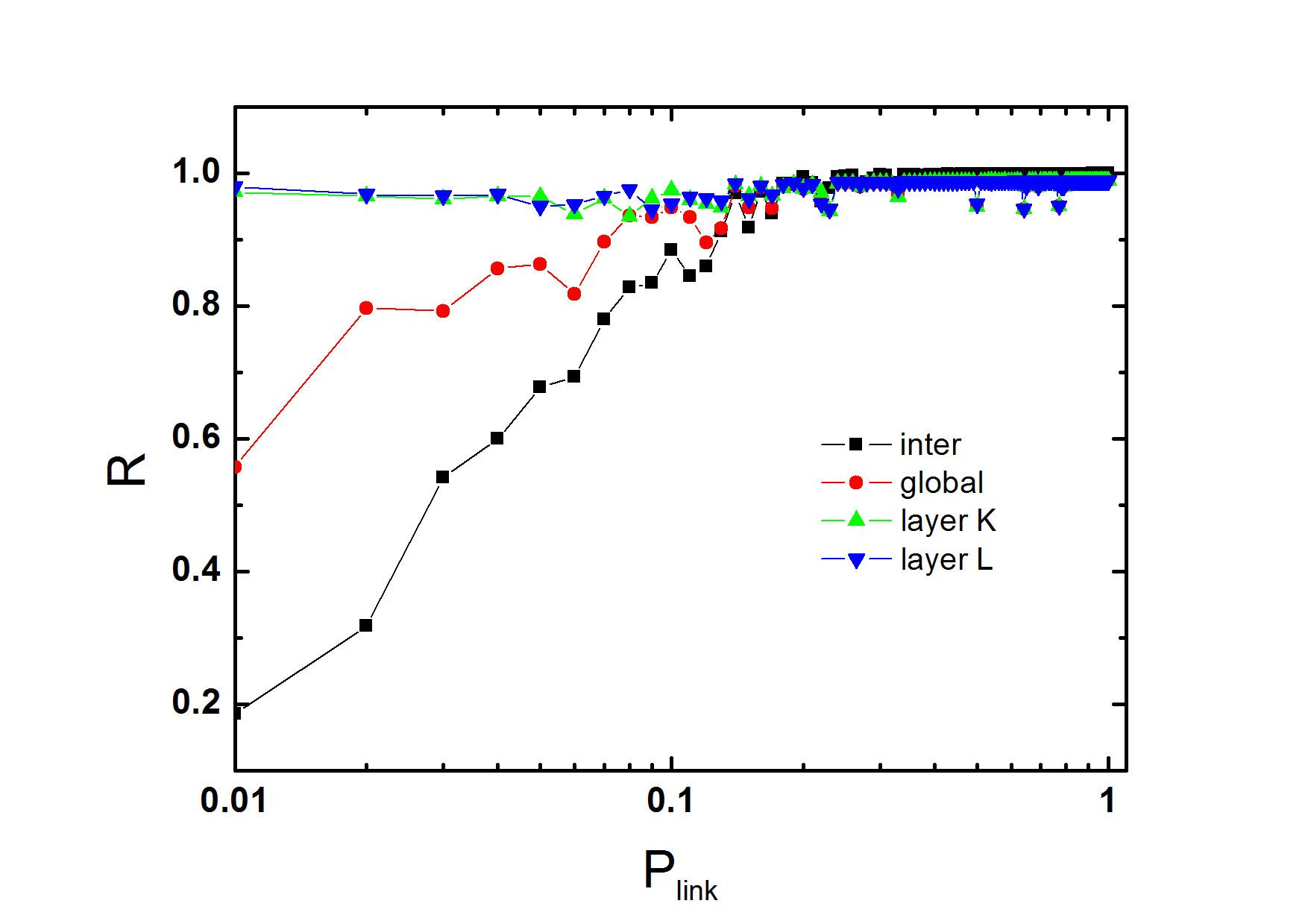}\\
     \caption{The inter-layer, intra-layer and global order parameter as a function of inter-link density, for \(\lambda_{inter}=\lambda_{intra}=5\). The intra-layer order parameters almost do not vary with \(P_{link}\). While for inter-layer and global order parameters, there exists a critical vale thresholding synchronization and non-synchronization. }\label{pvr}
\end{figure}

\section{\label{sec:level1}Conclusions}
In this letter, we study synchronization on single-layer and duplex spacial networks with total cost constraint. Fist of all, we investigate the influence of spacial exponent \(\alpha\) on synchronizability of single-layer Li networks. When \(\alpha>d+1\) \((d=2)\) , no matter what values of coupling strength are assigned, the networks can barely synchronize. For \(\alpha<d+1\) , the networks will change from non-synchronization to synchronization with increasing coupling strength. Then we study the effect of inter-layer and intra-layer coupling strength on synchronous behaviors for duplex spacial networks for \(\alpha=3\) with total cost constraint. We find that synchronization will emerge with strong enough inter-layer and intra-layer coupling. Furthermore we introduce order parameters for inter-layer, intra-layer and global synchronization to investigate synchronization processes. We note that when inter-layer coupling strength is larger than a certain value, no matter what value of intra-layer coupling strength is assigned, the inter-layer synchronization will always occur fist. And inter-layer interactions can improve the inter-layer and global synchronization. Interestingly, the synchronizability of a node is positively related with its degree, and for duplex networks this effect will be weakened by the inter-layer interactions. That is, inter-layer interactions play a crucial role in global synchronization for duplex Li networks. Finally, we study the impact of the inter-link density on global synchronization, and find that the existence of sparse inter-links can already lead to global synchronization of duplex Li networks. Those results for duplex Li networks can help us to manipulate the synchronization for a specific purpose, such as making two different groups reach an agreement in an optimal way.

\end{document}